\documentclass[floats,floatfix,showpacs,amssymb,prd,twocolumn,superscriptaddress,nofootinbib,nolongbibliography,reprint]{revtex4-1}

\usepackage{amssymb,amsmath,verbatim,mathtools,needspace,enumitem,etoolbox,graphicx,physics,microtype,afterpage,xspace,tabularx,lmodern,multirow}
\usepackage{gensymb}
\usepackage[dvipsnames, usenames]{xcolor}
\definecolor{linkcolor}{rgb}{0.0,0.3,0.5}
\usepackage[unicode, colorlinks=true, linkcolor=linkcolor, citecolor=linkcolor, filecolor=linkcolor, urlcolor=linkcolor, linktocpage, breaklinks]{hyperref}
\usepackage[all]{hypcap}
\usepackage[T1]{fontenc}
\usepackage[utf8]{inputenc}
\usepackage[usenames,dvipsnames]{xcolor}
\hypersetup{colorlinks=true,citecolor=romared,linkcolor=romared,urlcolor=romared}

\definecolor{romared}{RGB}{142,0,28}

\newcommand{\be}{\begin{equation}}
\newcommand{\ee}{\end{equation}}

\def\be{\begin{equation}}
\def\ee{\end{equation}}
\newcommand{\beq}{\begin{eqnarray}}
\newcommand{\eeq}{\end{eqnarray}}

\usepackage{makecell}
\usepackage{soul}

\newcolumntype{Y}{>{\centering\arraybackslash}X}

\begin{document}

\title{Destabilizing the Fundamental Mode of Black Holes: The Elephant and the Flea}

\author{Mark Ho-Yeuk Cheung}
\affiliation{Department of Physics and Astronomy, Johns Hopkins University, 3400 North Charles Street, Baltimore, Maryland, 21218, USA}
\author{Kyriakos Destounis}
\affiliation{Theoretical Astrophysics, IAAT, University of T{\"u}bingen, 72076 T{\"u}bingen, Germany}
\author{Rodrigo Panosso Macedo}
\affiliation{CENTRA, Departamento de F\'{\i}sica, Instituto Superior T\'ecnico--IST, Universidade de Lisboa--UL, Avenida Rovisco Pais 1, 1049 Lisboa, Portugal}
\affiliation{STAG Research Centre, University of Southampton, University Road SO17 1BJ, Southampton, United Kingdom}
\author{Emanuele Berti}
\affiliation{Department of Physics and Astronomy, Johns Hopkins University, 3400 North Charles Street, Baltimore, Maryland, 21218, USA}
\author{Vitor Cardoso} 
\affiliation{Niels Bohr International Academy, Niels Bohr Institute, Blegdamsvej 17, 2100 Copenhagen, Denmark}
\affiliation{CENTRA, Departamento de F\'{\i}sica, Instituto Superior T\'ecnico--IST, Universidade de Lisboa--UL, Avenida Rovisco Pais 1, 1049 Lisboa, Portugal}

\date{\today}

\begin{abstract}
  Recent work applying the notion of pseudospectrum to gravitational physics showed that the quasinormal mode spectrum of black holes is unstable, with the possible exception of the longest-lived (fundamental) mode.  
The fundamental mode dominates the expected signal in gravitational wave astronomy, and there is no reason why it should have privileged status.
We compute the quasinormal mode spectrum of two model problems where the Schwarzschild potential is perturbed by a small ``bump'' consisting of either a P\"oschl-Teller potential or a Gaussian, and we show that the fundamental mode is destabilized under generic perturbations. 
We present phase diagrams and study a simple double-barrier toy problem to clarify the conditions under which the spectral instability occurs.
\end{abstract}

\maketitle

\noindent
{\bf \em Introduction.} The advent of gravitational-wave (GW) astronomy~\cite{LIGOScientific:2016aoc,Abbott:2020niy} and of very long baseline interferometry~\cite{EventHorizonTelescope:2019dse,GRAVITY:2020gka} opened
exciting new windows to the invisible Universe.
Black holes (BHs) play a unique role in the endeavor to test our understanding of general relativity (GR) and in the search for new physics~\cite{Berti:2015itd,Barack:2018yly,Berti:2018cxi,Berti:2018vdi,Cardoso:2019rvt,Bertone:2018krk,Brito:2015oca}. 

According to the singularity theorems~\cite{Penrose:1964wq,Penrose:1969pc}, classical GR must fail in BH interiors.
Quantum mechanics in BH spacetimes also leads to puzzling consequences, such as the information paradox~\cite{Unruh:2017uaw,Mathur:2005zp,Giddings:2017mym}.
It is tempting to conjecture that a theory of quantum gravity will resolve these issues,
but the scale and nature of quantum gravity corrections to BH spacetimes is unknown. Uniqueness results in vacuum GR imply that BHs are the simplest macroscopic objects in the Universe~\cite{Chrusciel:2012jk}, and BHs do not ``polarize'' in binary systems~\cite{Binnington:2009bb,Damour:2009vw,Cardoso:2017cfl,LeTiec:2020bos,Chia:2020yla,Hui:2020xxx}.
The simplicity of BHs (whether isolated or in binaries) implies that they are ideal laboratories to probe the limitations of GR, as long as environmental effects or astrophysical uncertainties can be ignored. In this Letter we ask an important question: is it really possible to ignore environmental effects?

One of the tools to test the Kerr geometry is BH spectroscopy~\cite{Dreyer:2003bv,Berti:2005ys,Berti:2007zu}, now a thriving field~\cite{Baibhav:2017jhs,LIGOScientific:2016lio,Isi:2019aib,Carullo:2019flw,Laghi:2020rgl,Bustillo:2020buq,Capano:2021etf,Isi:2021iql}.
If a compact binary merger leads to the formation of a rotating BH, as predicted in GR, the spacetime should asymptote to the Kerr metric through a relaxation process during which it can be described as a perturbation of the Kerr metric. 
The late-time GW signal (the ``ringdown'') is a superposition of damped exponentials with complex frequencies known as the quasinormal modes (QNMs), which can be computed within perturbation theory as poles of the associated Green's function~\cite{Leaver:1986gd,Kokkotas:1999bd,Berti:2009kk}. The residues corresponding to these poles in the complex frequency plane dictate the amplitude of the response. To model a ringdown signal using Kerr QNM frequencies in vacuum, we should take into account the surrounding matter (even if it can be considered as a small perturbation). This is the main motivation of our work.

The behavior of the Green's function in the entire complex plane can be investigated using the mathematical notion of ``pseudospectrum''~\cite{1993Sci...261..578T,Jaramillo:2020tuu,Jaramillo:2021tmt,Gasperin:2021kfv,Destounis:2021lum}.
Through the pseudospectrum we can understand whether the QNM spectrum itself is stable under perturbations~\cite{Nollert:1996rf,Nollert:1998ys,Daghigh:2020jyk}. Recent work on the pseudospectrum showed that all Schwarzschild QNMs exhibit spectral instability, with the possible exception of the longest-lived (fundamental) mode~\cite{Jaramillo:2020tuu}. The fundamental QNM is expected to dominate the GW response of BHs, and its spectral stability is crucial for BH spectroscopy with GW observations~\cite{Jaramillo:2021tmt}.

In this Letter, we consider generic, small perturbations of the effective potential dictating the dynamics of GWs around Schwarzschild BHs consisting of tiny bumps, which may be produced, e.g., by matter in the local BH environment~\cite{Barausse:2014tra}, and we show that they inevitably lead to large shifts in the frequency and damping time of the fundamental mode.
The spectral instability of the fundamental mode has important implications for BH spectroscopy: while the overtone instability pointed out in Refs.~\cite{Jaramillo:2020tuu,Jaramillo:2021tmt} may not be easy to observe in the near future, the fundamental mode is already within the LIGO-Virgo detection range.

We will work in geometrical units ($G = c = 1$).

\begin{figure*}
	\includegraphics[width=\textwidth]{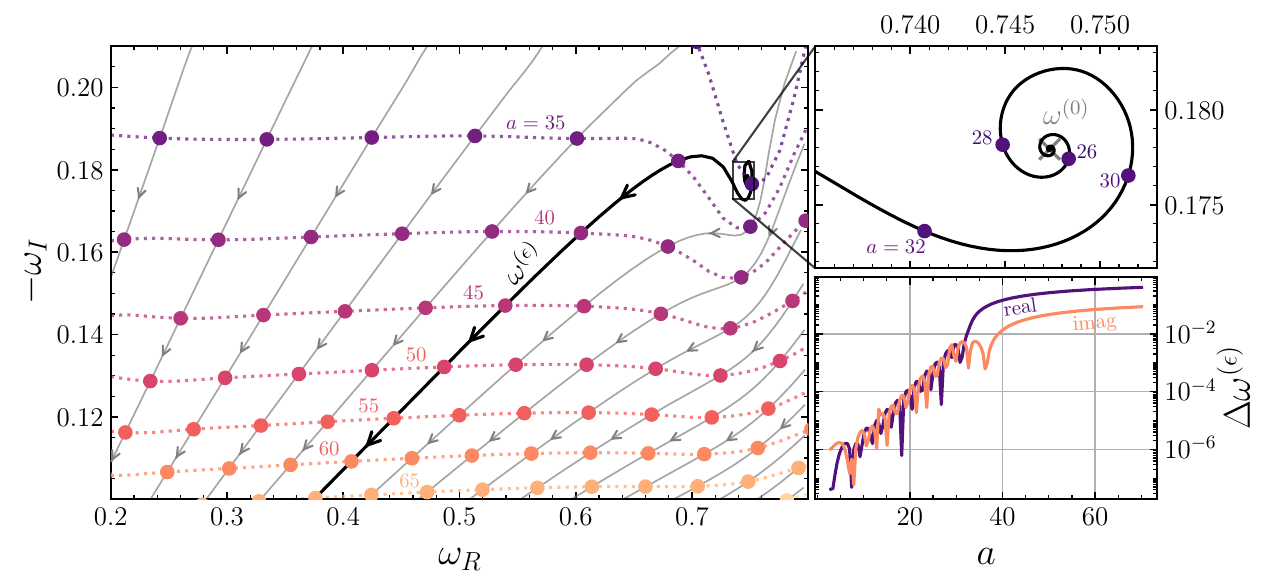}
	\caption{\label{fig:RWPT_main} Migration of $\ell=2$ QNMs as a function of the bump position $a$ for a P\"oschl-Teller perturbation with $\epsilon = 10^{-6}$.
	\textit{Left:} individual modes migrate along the black lines. The mode $\omega^{(\epsilon)}$ (bold line) reduces to the Schwarzschild fundamental mode when $\epsilon\to 0$, and the arrows indicate the direction of migration as $a$ increases. Modes with the same value of $a$ have the same color, and are connected with dotted lines. 
	\textit{Top right:} close-up view around the unperturbed fundamental QNM $\omega^{(0)}$.
	\textit{Bottom right:} real and imaginary parts of the migration distance $\Delta\omega^{(\epsilon)}$ of the perturbed fundamental QNM. We use units such that $2M=1$.
      }
\end{figure*}

\noindent
{\bf \em The Regge-Wheeler equation.}
Gravitational fluctuations in the background of a nonrotating BH with mass $M$ can be reduced to the study of a radial gauge-independent master function $\Psi$.
In Fourier space $\Psi$ obeys an ordinary differential equation~\cite{Regge:1957td,Chandrasekhar:1985kt}
\begin{equation}
\dfrac{d^2 \Psi}{d r_*^2} + \left[\omega^2 - V_{\ell}\right] \Psi = 0\,, \label{eq:RW}
\end{equation}
where the tortoise coordinate $r_*$ is defined in terms of the areal coordinate $r$ via $dr/dr_*=1-1/r$, $\omega$ is the Fourier variable, and
we set the Schwarzschild radius $2M=1$.
The angular coordinates were separated via an expansion in tensor spherical harmonics with angular number $\ell=2,3...$.
Without loss of generality we focus on odd-type gravitational perturbations, described by the Regge-Wheeler potential
\begin{equation}
V_{\ell}=\left(1-\frac{1}{r}\right)\left[\frac{\ell (\ell+1)}{r^2}-\frac{3}{r^3}\right]\,.\label{eq:RW_pot}
\end{equation}
Since GW emission is predominantly quadrupolar, we focus on the $\ell=2$ mode and write $V \equiv V_{2}$.
QNM frequencies are defined as the complex eigenvalues $\omega$ of Eq.~\eqref{eq:RW} such that perturbations are purely ingoing at the event horizon and outgoing at spatial infinity~\cite{Kokkotas:1999bd,Berti:2009kk}.
      
\noindent
{\bf \em Instability of the fundamental mode.} Consider now a small perturbation to the effective potential (induced e.g. by matter in the BH exterior~\cite{Barausse:2014tra}) of the form
\begin{equation} 
V_\epsilon=V+\epsilon \, V_{\rm bump}\,, \label{eq:replacement}
\end{equation} 
with $\epsilon \ll 1$ and $V_{\rm bump}$ a generic bump located at $r_* = a$, such that $V_{\rm bump}$ goes to zero as $r_*\to \pm \infty$ at least as fast as $V$.
Such a bump could be introduced by matter surrounding the BH (see~\cite{Chung:2021roh} or Supplemental Material for explicit examples).
We are interested in the complex QNM frequencies $\omega^{(\epsilon)}_{n}\equiv\omega^{(\epsilon)}_{n\, R}+i\omega^{(\epsilon)}_{n\, I}$ of the perturbed potential $V_\epsilon$. 
When the perturbation is added, the original fundamental mode $\omega_0^{(0)}$ migrates {\it continuously} in the complex plane along a curve which, in general, depends on $\epsilon$ and on the parameters characterizing $V_{\rm bump}$.  

Note that $\omega_0^{(\epsilon)}$ is not necessarily the fundamental mode of the perturbed potential $V_\epsilon$, which is defined as the QNM with the smallest $\left|\omega_I\right|$. We define $\varpi$ to be the fundamental mode of the perturbed potential $V_\epsilon$ and we set $\omega^{(\epsilon)} \equiv \omega^{(\epsilon)}_0$ -- i.e., we drop the subscript $0$ from the QNM frequencies that correspond to a continuous deformation of the original fundamental mode $\omega^{(0)}=\omega^{(0)}_0$.

\begin{figure*}
	\includegraphics[width=\textwidth]{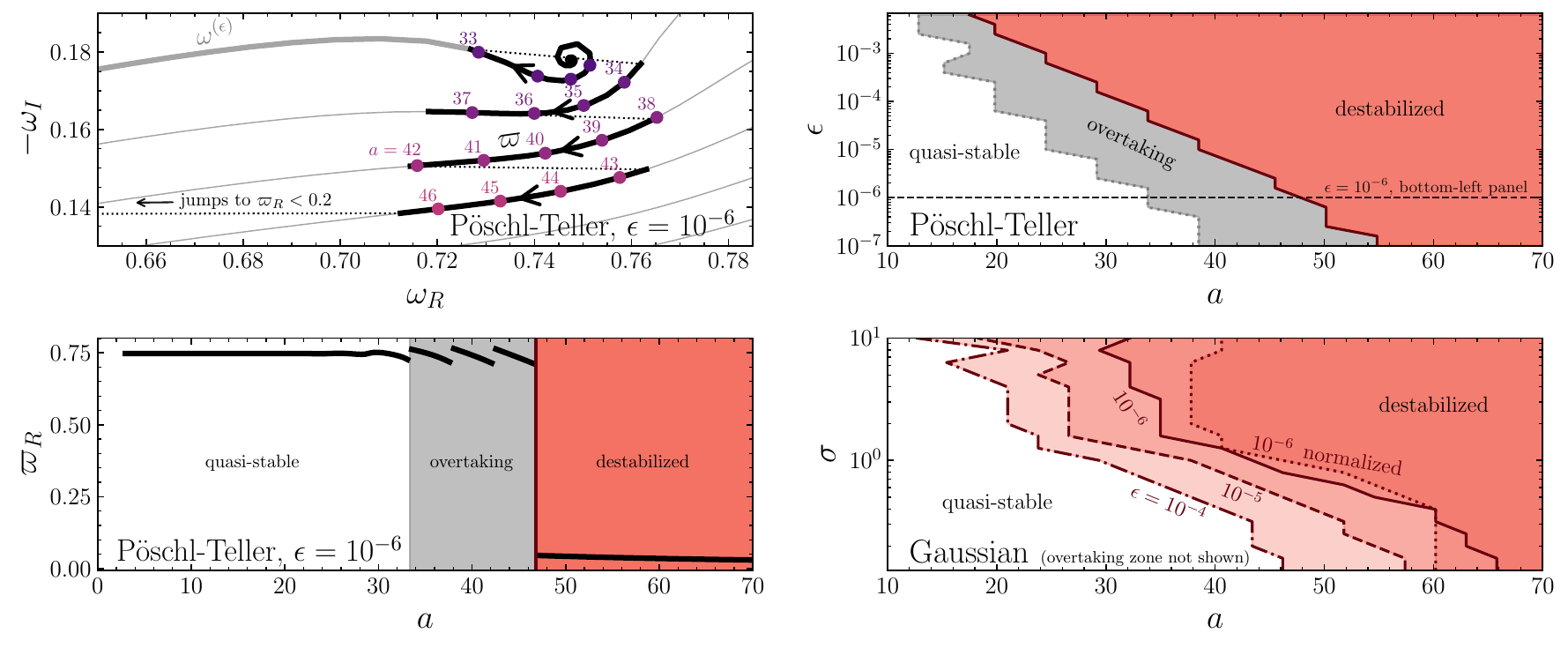}
	\caption{\label{fig:phase_diagrams} \textit{Top left:} migration of the fundamental mode $\varpi$ in the complex plane when $a$ is increased, for a P\"oschl-Teller bump with $\epsilon = 10^{-6}$.
	 Discontinuous jumps in $\varpi$ are shown with dotted lines. 
	 \textit{Bottom left:} variation of the real part of $\varpi$ in the top left panel as a function of $a$. 
	 \textit{Top right:} ``phase diagram'' of $\epsilon$ vs $a$ for a P\"oschl-Teller bump.
	  \textit{Bottom right:} ``phase diagram'' of $\sigma$ vs $a$ for a Gaussian bump with different values of $\epsilon$.
	}
\end{figure*}

We observe that one can destabilize the fundamental QNM in two different ways, as illustrated in Fig.~\ref{fig:RWPT_main}:

\noindent {\em (i) Destabilization via migration of the fundamental mode.} 
By measuring the variation $\Delta\omega^{(\epsilon)} = \omega^{(\epsilon)} - \omega^{(0)}$ in a continuous deformation of the original fundamental QNM frequency, for large enough $a$ we find regimes in which $\omega^{(\epsilon)} =\varpi$, but $|\Delta\omega^{(\epsilon)}/\omega^{(0)}| \gg \epsilon$. In this case the fundamental QNM is destabilized because it migrates over distances in the complex plane which are much larger than the scale of the perturbation $\epsilon$.

\noindent {\em (ii) Discontinuous overtaking of the fundamental mode.} In the regime $|\Delta\omega^{(\epsilon)}/\omega^{(0)}| \gg \epsilon$, QNMs which initially had large $|\omega_I|$ can ``overtake'' $\omega^{(\epsilon)}$ to become the new fundamental mode $\varpi$. Each overtaking causes a discontinuous jump in $\varpi_R$ which is orders of magnitude larger than $\epsilon$.

\noindent
{\bf \em P\"oschl-Teller and Gaussian bumps.} We demonstrate these phenomena by modeling
the perturbation $V_{\rm bump}$ either by the P\"oschl-Teller potential
\begin{equation}\label{eq:PTbump}
	V_{\rm PT}(r_* - a) = {\rm sech}^2\left(r_* - a\right),
\end{equation}
or by a Gaussian peak with varying width $\sigma$:
\begin{equation}\label{eq:Gaussian}
	V_{\rm G}(r_* - a) = \exp \left(\frac{-(r_* - a)^2}{2\sigma}\right).
\end{equation}
We use the shooting method to solve for the QNMs by integrating Eq.~\eqref{eq:RW} numerically from the boundaries to the center, and searching for the values of $\omega$ that give a matching solution (see e.g.~\cite{Pani:2013pma}). We have performed convergence tests and cross-checked our results against alternative numerical methods~\cite{Ansorg:2016ztf,Cardoso:2017soq}.

Figure~\ref{fig:RWPT_main} shows the modes for the P\"oschl-Teller bump as we increase $a$ for a fixed value of $\epsilon = 10^{-6}$. In the left panel we highlight in bold the curve traced by a continuous perturbation of the original fundamental mode. The top-right panel shows a close-up view into the trajectory near $\omega^{(0)}$ for moderate $a$, and the bottom-right panel shows the real and imaginary parts of the perturbed QNM frequency as functions of $a$. The perturbed QNM moves over regions such that $|\Delta \omega^{(\epsilon)}/\omega^{(0)}| \gg \epsilon$ for sufficient large $a$: the bottom-right panel shows that $|\Delta\omega^{(\epsilon)}|$ grows exponentially from $\sim10^{-6}$ to $\sim10^{-2}$ in the regime $10\lesssim a \lesssim 33$. This is the ``migration instability'' of item (i) above. 

The exponential growth with $a$ is related to the exponentially increasing nature of the eigenfunction $\Psi \sim e^{i\omega r_*}$: at large $r_*$, the response of the eigenfunction to a perturbative bump increases exponentially. In the mathematical literature, a similar exponential behavior is expected for small disturbances of symmetric multiwell potentials~\cite{1981CMaPh..80..223J,1984JPhA...17.2935G,Simon1985SemiclassicalAO,2019EJPh...40e5402S}. To the best of our knowledge, there are no analogous theorems for potentials of relevance to BH physics.

For $a \sim 30$, the arrows show that new modes move fast toward the bottom left of Fig.~\ref{fig:RWPT_main}. Eventually some of these modes overtake $\omega^{(\epsilon)}$: this is the (discontinuous) overtaking instability described in item (ii) above. 
For bumps at large enough distance $a$, the fundamental QNM can be destabilized by perturbations with $\epsilon\ll 1$. 

The top-left panel of Fig.~\ref{fig:phase_diagrams} shows the discontinuous overtaking instability of the fundamental mode in more detail. 
For a P\"oschl-Teller bump located at small values of $a$, the fundamental mode is still $\varpi=\omega^{(\epsilon)}$. Around $a \sim 33$, a new mode coming from the right overtakes $\omega^{(\epsilon)}$ and becomes the new fundamental mode $\varpi$, so that the real part of $\varpi$ has a discontinuity.
After three consecutive overtakings with $\Delta \varpi_R \sim O(10^{-2})$, $\varpi$ jumps one more time
to $\varpi_R < 0.2$, with $\Delta \varpi_R \sim O(10^{-1})$, and the fundamental QNM is completely destabilized.

The bottom-left panel of Fig.~\ref{fig:phase_diagrams} shows $\varpi_R$ as a function of $a$. We can identify three different regimes: a first regime where the fundamental mode is quasistable, a regime where multiple overtakings occur, and a third regime where the fundamental QNM is completely destabilized. In the latter regime, the separation between the real parts of two consecutive modes is given to a very good approximation by $\omega_{n+1, R} -\omega_{n, R} = \pi / a$: this is the expected characteristic behavior of modes trapped between two potential barriers located at distance $a$ from each other (see e.g.~\cite{Barausse:2014tra,Cardoso:2019rvt,Jaramillo:2021tmt}) and it can lead to multiple ringdown wave trains or ``echoes''~\cite{Cardoso:2016rao}.

We have repeated the analysis for P\"oschl-Teller perturbations with different amplitudes $\epsilon$.
The top-right panel of Fig.~\ref{fig:phase_diagrams} is a ``phase diagram'' in the $(a,\,\epsilon)$ plane showing where spectral instabilities are possible. The overtaking instability occurs as soon as we get into the gray area, while the top-right region corresponds to complete destabilization.  The bottom-left panel of Fig.~\ref{fig:phase_diagrams} is a cross section of this diagram, corresponding to the horizontal long-dashed line at $\epsilon=10^{-6}$. The trend is clear and consistent with the previous discussion: as $a$ increases, the values of $\epsilon$ needed to destabilize the spectrum decrease exponentially, as they should if the instability is indeed related to the exponentially increasing response of the wave function to the bump for large values of $a$. 

In the bottom-right panel of Fig.~\ref{fig:phase_diagrams} we show a similar ``phase diagram'' for Gaussian perturbations with different values of the amplitude $\epsilon$ and width $\sigma$. For clarity, in this case we show only the phase diagram boundaries corresponding to complete destabilization. A broader bump (i.e., a bump with larger values of $\sigma$) is more effective at destabilizing the fundamental mode.
This is not simply due to the fact that we are fixing $\epsilon$ and increasing $\sigma$, thus producing a ``stronger'' perturbation of the original potential.
We have repeated the analysis normalizing the Gaussian bump by $1 / \sqrt{2\pi \sigma}$, and we obtain qualitatively similar results (see the dotted line in the bottom-right panel). The fact that the QNM instability occurs for smaller $a$ when the bump is wider is not an artifact of the larger area under the curve. 
As we show in Supplemental Material, the qualitative features of this study are confirmed by the analysis of a simple toy model consisting of a double rectangular barrier (cf.~\cite{Barausse:2014tra}).

\noindent
{\bf \em Conclusions.} We have studied two model problems in which the potential describing gravitational perturbations of a Schwarzschild BH is perturbed by either a P\"oschl-Teller or a Gaussian bump of amplitude $\epsilon$ located at distance $\approx a$ from the light ring. We have demonstrated that the fundamental mode of the Schwarzschild potential can be destabilized in two ways: either because it
migrates continuously by an amount $|\Delta\omega^{(\epsilon)}/\omega^{(0)}|\gg \epsilon$ when the perturbing bump is located at large enough $a$ (``migration instability''), or because of the appearance of a new family of ``trapped modes'' in between the two potential barriers that can overtake the original fundamental mode (``overtaking instability''). We have shown through ``instability phase diagrams'' that the value of $\epsilon$ needed to destabilize the spectrum decreases exponentially as $a$ increases, and that broad bumps are more likely to destabilize the fundamental mode. The analysis is therefore consistent with the conclusions of Ref.~\cite{Jaramillo:2020tuu}: short length-scale (ultraviolet) perturbations do not destabilize the fundamental QNM, but large-scale (infrared) perturbations might.

How does spin affect the instability, and does the instability play a role in gravitational turbulence for near-extremal Kerr BHs~\cite{Yang:2014tla}?
What are the implications of our results for modeling the ringdown of BHs surrounded by matter or other forms of ``hair''?
Is this instability a threat to the BH spectroscopy program in GW astronomy~\cite{Jaramillo:2021tmt}, and can it circumvent the failure of determinism in GR~\cite{Cardoso:2017soq,Cardoso:2018nvb}? More fundamentally, do infrared and/or ultraviolet corrections to general relativity affect at a fundamental level the meaning of the QNM spectrum and BH stability? These are important questions that must be addressed through numerical simulations and further theoretical work (see Ref.~\cite{Cardoso:2021wlq} for first steps in this direction).

We thank Jos\'e Luis Jaramillo for pointing out
Refs.~\cite{1981CMaPh..80..223J,1984JPhA...17.2935G,Simon1985SemiclassicalAO}
about the ``flea on the elephant'' effect, as well as Luca Reali,
Thomas Helfer, and the other members of the JHU gravity group for
useful discussions.
M.H.Y.C. and E.B. are supported by NSF Grants No. PHY-1912550 and AST-2006538, NASA ATP Grants No. 17-ATP17-0225 and 19-ATP19-0051, NSF-XSEDE Grant No. PHY-090003, and NSF Grant PHY-20043. 
V.C. is a Villum Investigator supported by VILLUM FONDEN (Grant No. 37766).
V.C. acknowledges financial support provided under the European Union's H2020 ERC 
Consolidator Grant ``Matter and strong-field gravity: New frontiers in Einstein's 
theory'' Grant No. MaGRaTh--646597.
R.P.M acknowledges financial support provided by COST Action CA16104 via the Short Term Scientific Mission grant and from STFC via Grant No. ST/V000551/1.
This project has received funding from the European Union's Horizon 2020 research and innovation programme under the Marie Sklodowska-Curie Grant No. 101007855.
We thank FCT for financial support through Project~No.~UIDB/00099/2020.
We acknowledge financial support provided by FCT/Portugal through grants PTDC/MAT-APL/30043/2017 and PTDC/FIS-AST/7002/2020.
This research project was conducted using computational resources at the Maryland Advanced Research Computing Center (MARCC).
The authors acknowledge the Texas Advanced Computing Center (TACC) at The
University of Texas at Austin for providing HPC resources that have contributed
to the research results reported within this Letter.\footnote{\url{http://www.tacc.utexas.edu}}

\bibliography{QNMinstabilityBib}

\begin{thebibliography}{61}%
\makeatletter
\providecommand \@ifxundefined [1]{%
 \@ifx{#1\undefined}
}%
\providecommand \@ifnum [1]{%
 \ifnum #1\expandafter \@firstoftwo
 \else \expandafter \@secondoftwo
 \fi
}%
\providecommand \@ifx [1]{%
 \ifx #1\expandafter \@firstoftwo
 \else \expandafter \@secondoftwo
 \fi
}%
\providecommand \natexlab [1]{#1}%
\providecommand \enquote  [1]{``#1''}%
\providecommand \bibnamefont  [1]{#1}%
\providecommand \bibfnamefont [1]{#1}%
\providecommand \citenamefont [1]{#1}%
\providecommand \href@noop [0]{\@secondoftwo}%
\providecommand \href [0]{\begingroup \@sanitize@url \@href}%
\providecommand \@href[1]{\@@startlink{#1}\@@href}%
\providecommand \@@href[1]{\endgroup#1\@@endlink}%
\providecommand \@sanitize@url [0]{\catcode `\\12\catcode `\$12\catcode
  `\&12\catcode `\#12\catcode `\^12\catcode `\_12\catcode `\%12\relax}%
\providecommand \@@startlink[1]{}%
\providecommand \@@endlink[0]{}%
\providecommand \url  [0]{\begingroup\@sanitize@url \@url }%
\providecommand \@url [1]{\endgroup\@href {#1}{\urlprefix }}%
\providecommand \urlprefix  [0]{URL }%
\providecommand \Eprint [0]{\href }%
\providecommand \doibase [0]{http://dx.doi.org/}%
\providecommand \selectlanguage [0]{\@gobble}%
\providecommand \bibinfo  [0]{\@secondoftwo}%
\providecommand \bibfield  [0]{\@secondoftwo}%
\providecommand \translation [1]{[#1]}%
\providecommand \BibitemOpen [0]{}%
\providecommand \bibitemStop [0]{}%
\providecommand \bibitemNoStop [0]{.\EOS\space}%
\providecommand \EOS [0]{\spacefactor3000\relax}%
\providecommand \BibitemShut  [1]{\csname bibitem#1\endcsname}%
\let\auto@bib@innerbib\@empty
\bibitem [{\citenamefont {Abbott}\ \emph
  {et~al.}(2016{\natexlab{a}})\citenamefont {Abbott} \emph
  {et~al.}}]{LIGOScientific:2016aoc}%
  \BibitemOpen
  \bibfield  {author} {\bibinfo {author} {\bibfnamefont {B.~P.}\ \bibnamefont
  {Abbott}} \emph {et~al.} (\bibinfo {collaboration} {LIGO Scientific,
  Virgo}),\ }\href {\doibase 10.1103/PhysRevLett.116.061102} {\bibfield
  {journal} {\bibinfo  {journal} {Phys. Rev. Lett.}\ }\textbf {\bibinfo
  {volume} {116}},\ \bibinfo {pages} {061102} (\bibinfo {year}
  {2016}{\natexlab{a}})},\ \Eprint {http://arxiv.org/abs/1602.03837}
  {arXiv:1602.03837 [gr-qc]} \BibitemShut {NoStop}%
\bibitem [{\citenamefont {Abbott}\ \emph {et~al.}(2021)\citenamefont {Abbott}
  \emph {et~al.}}]{Abbott:2020niy}%
  \BibitemOpen
  \bibfield  {author} {\bibinfo {author} {\bibfnamefont {R.}~\bibnamefont
  {Abbott}} \emph {et~al.} (\bibinfo {collaboration} {LIGO Scientific,
  Virgo}),\ }\href {\doibase 10.1103/PhysRevX.11.021053} {\bibfield  {journal}
  {\bibinfo  {journal} {Phys. Rev. X}\ }\textbf {\bibinfo {volume} {11}},\
  \bibinfo {pages} {021053} (\bibinfo {year} {2021})},\ \Eprint
  {http://arxiv.org/abs/2010.14527} {arXiv:2010.14527 [gr-qc]} \BibitemShut
  {NoStop}%
\bibitem [{\citenamefont {Akiyama}\ \emph {et~al.}(2019)\citenamefont {Akiyama}
  \emph {et~al.}}]{EventHorizonTelescope:2019dse}%
  \BibitemOpen
  \bibfield  {author} {\bibinfo {author} {\bibfnamefont {K.}~\bibnamefont
  {Akiyama}} \emph {et~al.} (\bibinfo {collaboration} {Event Horizon
  Telescope}),\ }\href {\doibase 10.3847/2041-8213/ab0ec7} {\bibfield
  {journal} {\bibinfo  {journal} {Astrophys. J. Lett.}\ }\textbf {\bibinfo
  {volume} {875}},\ \bibinfo {pages} {L1} (\bibinfo {year} {2019})},\ \Eprint
  {http://arxiv.org/abs/1906.11238} {arXiv:1906.11238 [astro-ph.GA]}
  \BibitemShut {NoStop}%
\bibitem [{\citenamefont {Abuter}\ \emph {et~al.}(2020)\citenamefont {Abuter}
  \emph {et~al.}}]{GRAVITY:2020gka}%
  \BibitemOpen
  \bibfield  {author} {\bibinfo {author} {\bibfnamefont {R.}~\bibnamefont
  {Abuter}} \emph {et~al.} (\bibinfo {collaboration} {GRAVITY}),\ }\href
  {\doibase 10.1051/0004-6361/202037813} {\bibfield  {journal} {\bibinfo
  {journal} {Astron. Astrophys.}\ }\textbf {\bibinfo {volume} {636}},\ \bibinfo
  {pages} {L5} (\bibinfo {year} {2020})},\ \Eprint
  {http://arxiv.org/abs/2004.07187} {arXiv:2004.07187 [astro-ph.GA]}
  \BibitemShut {NoStop}%
\bibitem [{\citenamefont {Berti}\ \emph {et~al.}(2015)\citenamefont {Berti}
  \emph {et~al.}}]{Berti:2015itd}%
  \BibitemOpen
  \bibfield  {author} {\bibinfo {author} {\bibfnamefont {E.}~\bibnamefont
  {Berti}} \emph {et~al.},\ }\href {\doibase 10.1088/0264-9381/32/24/243001}
  {\bibfield  {journal} {\bibinfo  {journal} {Class. Quant. Grav.}\ }\textbf
  {\bibinfo {volume} {32}},\ \bibinfo {pages} {243001} (\bibinfo {year}
  {2015})},\ \Eprint {http://arxiv.org/abs/1501.07274} {arXiv:1501.07274
  [gr-qc]} \BibitemShut {NoStop}%
\bibitem [{\citenamefont {Barack}\ \emph {et~al.}(2019)\citenamefont {Barack}
  \emph {et~al.}}]{Barack:2018yly}%
  \BibitemOpen
  \bibfield  {author} {\bibinfo {author} {\bibfnamefont {L.}~\bibnamefont
  {Barack}} \emph {et~al.},\ }\href {\doibase 10.1088/1361-6382/ab0587}
  {\bibfield  {journal} {\bibinfo  {journal} {Class. Quant. Grav.}\ }\textbf
  {\bibinfo {volume} {36}},\ \bibinfo {pages} {143001} (\bibinfo {year}
  {2019})},\ \Eprint {http://arxiv.org/abs/1806.05195} {arXiv:1806.05195
  [gr-qc]} \BibitemShut {NoStop}%
\bibitem [{\citenamefont {Berti}\ \emph
  {et~al.}(2018{\natexlab{a}})\citenamefont {Berti}, \citenamefont {Yagi},\
  and\ \citenamefont {Yunes}}]{Berti:2018cxi}%
  \BibitemOpen
  \bibfield  {author} {\bibinfo {author} {\bibfnamefont {E.}~\bibnamefont
  {Berti}}, \bibinfo {author} {\bibfnamefont {K.}~\bibnamefont {Yagi}}, \ and\
  \bibinfo {author} {\bibfnamefont {N.}~\bibnamefont {Yunes}},\ }\href
  {\doibase 10.1007/s10714-018-2362-8} {\bibfield  {journal} {\bibinfo
  {journal} {Gen. Rel. Grav.}\ }\textbf {\bibinfo {volume} {50}},\ \bibinfo
  {pages} {46} (\bibinfo {year} {2018}{\natexlab{a}})},\ \Eprint
  {http://arxiv.org/abs/1801.03208} {arXiv:1801.03208 [gr-qc]} \BibitemShut
  {NoStop}%
\bibitem [{\citenamefont {Berti}\ \emph
  {et~al.}(2018{\natexlab{b}})\citenamefont {Berti}, \citenamefont {Yagi},
  \citenamefont {Yang},\ and\ \citenamefont {Yunes}}]{Berti:2018vdi}%
  \BibitemOpen
  \bibfield  {author} {\bibinfo {author} {\bibfnamefont {E.}~\bibnamefont
  {Berti}}, \bibinfo {author} {\bibfnamefont {K.}~\bibnamefont {Yagi}},
  \bibinfo {author} {\bibfnamefont {H.}~\bibnamefont {Yang}}, \ and\ \bibinfo
  {author} {\bibfnamefont {N.}~\bibnamefont {Yunes}},\ }\href {\doibase
  10.1007/s10714-018-2372-6} {\bibfield  {journal} {\bibinfo  {journal} {Gen.
  Rel. Grav.}\ }\textbf {\bibinfo {volume} {50}},\ \bibinfo {pages} {49}
  (\bibinfo {year} {2018}{\natexlab{b}})},\ \Eprint
  {http://arxiv.org/abs/1801.03587} {arXiv:1801.03587 [gr-qc]} \BibitemShut
  {NoStop}%
\bibitem [{\citenamefont {Cardoso}\ and\ \citenamefont
  {Pani}(2019)}]{Cardoso:2019rvt}%
  \BibitemOpen
  \bibfield  {author} {\bibinfo {author} {\bibfnamefont {V.}~\bibnamefont
  {Cardoso}}\ and\ \bibinfo {author} {\bibfnamefont {P.}~\bibnamefont {Pani}},\
  }\href {\doibase 10.1007/s41114-019-0020-4} {\bibfield  {journal} {\bibinfo
  {journal} {Living Rev. Rel.}\ }\textbf {\bibinfo {volume} {22}},\ \bibinfo
  {pages} {4} (\bibinfo {year} {2019})},\ \Eprint
  {http://arxiv.org/abs/1904.05363} {arXiv:1904.05363 [gr-qc]} \BibitemShut
  {NoStop}%
\bibitem [{\citenamefont {Bertone}\ and\ \citenamefont
  {Tait}(2018)}]{Bertone:2018krk}%
  \BibitemOpen
  \bibfield  {author} {\bibinfo {author} {\bibfnamefont {G.}~\bibnamefont
  {Bertone}}\ and\ \bibinfo {author} {\bibfnamefont {T.}~\bibnamefont {Tait},
  \bibfnamefont {M.~P.}},\ }\href {\doibase 10.1038/s41586-018-0542-z}
  {\bibfield  {journal} {\bibinfo  {journal} {Nature}\ }\textbf {\bibinfo
  {volume} {562}},\ \bibinfo {pages} {51} (\bibinfo {year} {2018})},\ \Eprint
  {http://arxiv.org/abs/1810.01668} {arXiv:1810.01668 [astro-ph.CO]}
  \BibitemShut {NoStop}%
\bibitem [{\citenamefont {Brito}\ \emph {et~al.}(2015)\citenamefont {Brito},
  \citenamefont {Cardoso},\ and\ \citenamefont {Pani}}]{Brito:2015oca}%
  \BibitemOpen
  \bibfield  {author} {\bibinfo {author} {\bibfnamefont {R.}~\bibnamefont
  {Brito}}, \bibinfo {author} {\bibfnamefont {V.}~\bibnamefont {Cardoso}}, \
  and\ \bibinfo {author} {\bibfnamefont {P.}~\bibnamefont {Pani}},\ }\href
  {\doibase 10.1007/978-3-319-19000-6} {\bibfield  {journal} {\bibinfo
  {journal} {Lect. Notes Phys.}\ }\textbf {\bibinfo {volume} {906}},\ \bibinfo
  {pages} {pp.1} (\bibinfo {year} {2015})},\ \Eprint
  {http://arxiv.org/abs/1501.06570} {arXiv:1501.06570 [gr-qc]} \BibitemShut
  {NoStop}%
\bibitem [{\citenamefont {Penrose}(1965)}]{Penrose:1964wq}%
  \BibitemOpen
  \bibfield  {author} {\bibinfo {author} {\bibfnamefont {R.}~\bibnamefont
  {Penrose}},\ }\href {\doibase 10.1103/PhysRevLett.14.57} {\bibfield
  {journal} {\bibinfo  {journal} {Phys. Rev. Lett.}\ }\textbf {\bibinfo
  {volume} {14}},\ \bibinfo {pages} {57} (\bibinfo {year} {1965})}\BibitemShut
  {NoStop}%
\bibitem [{\citenamefont {Penrose}(1969)}]{Penrose:1969pc}%
  \BibitemOpen
  \bibfield  {author} {\bibinfo {author} {\bibfnamefont {R.}~\bibnamefont
  {Penrose}},\ }\href {\doibase 10.1023/A:1016578408204} {\bibfield  {journal}
  {\bibinfo  {journal} {Riv. Nuovo Cim.}\ }\textbf {\bibinfo {volume} {1}},\
  \bibinfo {pages} {252} (\bibinfo {year} {1969})}\BibitemShut {NoStop}%
\bibitem [{\citenamefont {Unruh}\ and\ \citenamefont
  {Wald}(2017)}]{Unruh:2017uaw}%
  \BibitemOpen
  \bibfield  {author} {\bibinfo {author} {\bibfnamefont {W.~G.}\ \bibnamefont
  {Unruh}}\ and\ \bibinfo {author} {\bibfnamefont {R.~M.}\ \bibnamefont
  {Wald}},\ }\href {\doibase 10.1088/1361-6633/aa778e} {\bibfield  {journal}
  {\bibinfo  {journal} {Rept. Prog. Phys.}\ }\textbf {\bibinfo {volume} {80}},\
  \bibinfo {pages} {092002} (\bibinfo {year} {2017})},\ \Eprint
  {http://arxiv.org/abs/1703.02140} {arXiv:1703.02140 [hep-th]} \BibitemShut
  {NoStop}%
\bibitem [{\citenamefont {Mathur}(2005)}]{Mathur:2005zp}%
  \BibitemOpen
  \bibfield  {author} {\bibinfo {author} {\bibfnamefont {S.~D.}\ \bibnamefont
  {Mathur}},\ }\href {\doibase 10.1002/prop.200410203} {\bibfield  {journal}
  {\bibinfo  {journal} {Fortsch. Phys.}\ }\textbf {\bibinfo {volume} {53}},\
  \bibinfo {pages} {793} (\bibinfo {year} {2005})},\ \Eprint
  {http://arxiv.org/abs/hep-th/0502050} {arXiv:hep-th/0502050} \BibitemShut
  {NoStop}%
\bibitem [{\citenamefont {Giddings}(2017)}]{Giddings:2017mym}%
  \BibitemOpen
  \bibfield  {author} {\bibinfo {author} {\bibfnamefont {S.~B.}\ \bibnamefont
  {Giddings}},\ }\href {\doibase 10.1007/JHEP12(2017)047} {\bibfield  {journal}
  {\bibinfo  {journal} {JHEP}\ }\textbf {\bibinfo {volume} {12}},\ \bibinfo
  {pages} {047} (\bibinfo {year} {2017})},\ \Eprint
  {http://arxiv.org/abs/1701.08765} {arXiv:1701.08765 [hep-th]} \BibitemShut
  {NoStop}%
\bibitem [{\citenamefont {Chrusciel}\ \emph {et~al.}(2012)\citenamefont
  {Chrusciel}, \citenamefont {Lopes~Costa},\ and\ \citenamefont
  {Heusler}}]{Chrusciel:2012jk}%
  \BibitemOpen
  \bibfield  {author} {\bibinfo {author} {\bibfnamefont {P.~T.}\ \bibnamefont
  {Chrusciel}}, \bibinfo {author} {\bibfnamefont {J.}~\bibnamefont
  {Lopes~Costa}}, \ and\ \bibinfo {author} {\bibfnamefont {M.}~\bibnamefont
  {Heusler}},\ }\href {\doibase 10.12942/lrr-2012-7} {\bibfield  {journal}
  {\bibinfo  {journal} {Living Rev. Rel.}\ }\textbf {\bibinfo {volume} {15}},\
  \bibinfo {pages} {7} (\bibinfo {year} {2012})},\ \Eprint
  {http://arxiv.org/abs/1205.6112} {arXiv:1205.6112 [gr-qc]} \BibitemShut
  {NoStop}%
\bibitem [{\citenamefont {Binnington}\ and\ \citenamefont
  {Poisson}(2009)}]{Binnington:2009bb}%
  \BibitemOpen
  \bibfield  {author} {\bibinfo {author} {\bibfnamefont {T.}~\bibnamefont
  {Binnington}}\ and\ \bibinfo {author} {\bibfnamefont {E.}~\bibnamefont
  {Poisson}},\ }\href {\doibase 10.1103/PhysRevD.80.084018} {\bibfield
  {journal} {\bibinfo  {journal} {Phys. Rev. D}\ }\textbf {\bibinfo {volume}
  {80}},\ \bibinfo {pages} {084018} (\bibinfo {year} {2009})},\ \Eprint
  {http://arxiv.org/abs/0906.1366} {arXiv:0906.1366 [gr-qc]} \BibitemShut
  {NoStop}%
\bibitem [{\citenamefont {Damour}\ and\ \citenamefont
  {Nagar}(2009)}]{Damour:2009vw}%
  \BibitemOpen
  \bibfield  {author} {\bibinfo {author} {\bibfnamefont {T.}~\bibnamefont
  {Damour}}\ and\ \bibinfo {author} {\bibfnamefont {A.}~\bibnamefont {Nagar}},\
  }\href {\doibase 10.1103/PhysRevD.80.084035} {\bibfield  {journal} {\bibinfo
  {journal} {Phys. Rev. D}\ }\textbf {\bibinfo {volume} {80}},\ \bibinfo
  {pages} {084035} (\bibinfo {year} {2009})},\ \Eprint
  {http://arxiv.org/abs/0906.0096} {arXiv:0906.0096 [gr-qc]} \BibitemShut
  {NoStop}%
\bibitem [{\citenamefont {Cardoso}\ \emph {et~al.}(2017)\citenamefont
  {Cardoso}, \citenamefont {Franzin}, \citenamefont {Maselli}, \citenamefont
  {Pani},\ and\ \citenamefont {Raposo}}]{Cardoso:2017cfl}%
  \BibitemOpen
  \bibfield  {author} {\bibinfo {author} {\bibfnamefont {V.}~\bibnamefont
  {Cardoso}}, \bibinfo {author} {\bibfnamefont {E.}~\bibnamefont {Franzin}},
  \bibinfo {author} {\bibfnamefont {A.}~\bibnamefont {Maselli}}, \bibinfo
  {author} {\bibfnamefont {P.}~\bibnamefont {Pani}}, \ and\ \bibinfo {author}
  {\bibfnamefont {G.}~\bibnamefont {Raposo}},\ }\href {\doibase
  10.1103/PhysRevD.95.084014} {\bibfield  {journal} {\bibinfo  {journal} {Phys.
  Rev. D}\ }\textbf {\bibinfo {volume} {95}},\ \bibinfo {pages} {084014}
  (\bibinfo {year} {2017})},\ \bibinfo {note} {[Addendum: Phys.Rev.D 95, 089901
  (2017)]},\ \Eprint {http://arxiv.org/abs/1701.01116} {arXiv:1701.01116
  [gr-qc]} \BibitemShut {NoStop}%
\bibitem [{\citenamefont {Le~Tiec}\ \emph {et~al.}(2021)\citenamefont
  {Le~Tiec}, \citenamefont {Casals},\ and\ \citenamefont
  {Franzin}}]{LeTiec:2020bos}%
  \BibitemOpen
  \bibfield  {author} {\bibinfo {author} {\bibfnamefont {A.}~\bibnamefont
  {Le~Tiec}}, \bibinfo {author} {\bibfnamefont {M.}~\bibnamefont {Casals}}, \
  and\ \bibinfo {author} {\bibfnamefont {E.}~\bibnamefont {Franzin}},\ }\href
  {\doibase 10.1103/PhysRevD.103.084021} {\bibfield  {journal} {\bibinfo
  {journal} {Phys. Rev. D}\ }\textbf {\bibinfo {volume} {103}},\ \bibinfo
  {pages} {084021} (\bibinfo {year} {2021})},\ \Eprint
  {http://arxiv.org/abs/2010.15795} {arXiv:2010.15795 [gr-qc]} \BibitemShut
  {NoStop}%
\bibitem [{\citenamefont {Chia}(2021)}]{Chia:2020yla}%
  \BibitemOpen
  \bibfield  {author} {\bibinfo {author} {\bibfnamefont {H.~S.}\ \bibnamefont
  {Chia}},\ }\href {\doibase 10.1103/PhysRevD.104.024013} {\bibfield  {journal}
  {\bibinfo  {journal} {Phys. Rev. D}\ }\textbf {\bibinfo {volume} {104}},\
  \bibinfo {pages} {024013} (\bibinfo {year} {2021})},\ \Eprint
  {http://arxiv.org/abs/2010.07300} {arXiv:2010.07300 [gr-qc]} \BibitemShut
  {NoStop}%
\bibitem [{\citenamefont {Hui}\ \emph {et~al.}(2021)\citenamefont {Hui},
  \citenamefont {Joyce}, \citenamefont {Penco}, \citenamefont {Santoni},\ and\
  \citenamefont {Solomon}}]{Hui:2020xxx}%
  \BibitemOpen
  \bibfield  {author} {\bibinfo {author} {\bibfnamefont {L.}~\bibnamefont
  {Hui}}, \bibinfo {author} {\bibfnamefont {A.}~\bibnamefont {Joyce}}, \bibinfo
  {author} {\bibfnamefont {R.}~\bibnamefont {Penco}}, \bibinfo {author}
  {\bibfnamefont {L.}~\bibnamefont {Santoni}}, \ and\ \bibinfo {author}
  {\bibfnamefont {A.~R.}\ \bibnamefont {Solomon}},\ }\href {\doibase
  10.1088/1475-7516/2021/04/052} {\bibfield  {journal} {\bibinfo  {journal}
  {JCAP}\ }\textbf {\bibinfo {volume} {04}},\ \bibinfo {pages} {052} (\bibinfo
  {year} {2021})},\ \Eprint {http://arxiv.org/abs/2010.00593} {arXiv:2010.00593
  [hep-th]} \BibitemShut {NoStop}%
\bibitem [{\citenamefont {Dreyer}\ \emph {et~al.}(2004)\citenamefont {Dreyer},
  \citenamefont {Kelly}, \citenamefont {Krishnan}, \citenamefont {Finn},
  \citenamefont {Garrison},\ and\ \citenamefont
  {Lopez-Aleman}}]{Dreyer:2003bv}%
  \BibitemOpen
  \bibfield  {author} {\bibinfo {author} {\bibfnamefont {O.}~\bibnamefont
  {Dreyer}}, \bibinfo {author} {\bibfnamefont {B.~J.}\ \bibnamefont {Kelly}},
  \bibinfo {author} {\bibfnamefont {B.}~\bibnamefont {Krishnan}}, \bibinfo
  {author} {\bibfnamefont {L.~S.}\ \bibnamefont {Finn}}, \bibinfo {author}
  {\bibfnamefont {D.}~\bibnamefont {Garrison}}, \ and\ \bibinfo {author}
  {\bibfnamefont {R.}~\bibnamefont {Lopez-Aleman}},\ }\href {\doibase
  10.1088/0264-9381/21/4/003} {\bibfield  {journal} {\bibinfo  {journal}
  {Class. Quant. Grav.}\ }\textbf {\bibinfo {volume} {21}},\ \bibinfo {pages}
  {787} (\bibinfo {year} {2004})},\ \Eprint
  {http://arxiv.org/abs/gr-qc/0309007} {arXiv:gr-qc/0309007} \BibitemShut
  {NoStop}%
\bibitem [{\citenamefont {Berti}\ \emph {et~al.}(2006)\citenamefont {Berti},
  \citenamefont {Cardoso},\ and\ \citenamefont {Will}}]{Berti:2005ys}%
  \BibitemOpen
  \bibfield  {author} {\bibinfo {author} {\bibfnamefont {E.}~\bibnamefont
  {Berti}}, \bibinfo {author} {\bibfnamefont {V.}~\bibnamefont {Cardoso}}, \
  and\ \bibinfo {author} {\bibfnamefont {C.~M.}\ \bibnamefont {Will}},\ }\href
  {\doibase 10.1103/PhysRevD.73.064030} {\bibfield  {journal} {\bibinfo
  {journal} {Phys. Rev. D}\ }\textbf {\bibinfo {volume} {73}},\ \bibinfo
  {pages} {064030} (\bibinfo {year} {2006})},\ \Eprint
  {http://arxiv.org/abs/gr-qc/0512160} {arXiv:gr-qc/0512160} \BibitemShut
  {NoStop}%
\bibitem [{\citenamefont {Berti}\ \emph {et~al.}(2007)\citenamefont {Berti},
  \citenamefont {Cardoso}, \citenamefont {Cardoso},\ and\ \citenamefont
  {Cavaglia}}]{Berti:2007zu}%
  \BibitemOpen
  \bibfield  {author} {\bibinfo {author} {\bibfnamefont {E.}~\bibnamefont
  {Berti}}, \bibinfo {author} {\bibfnamefont {J.}~\bibnamefont {Cardoso}},
  \bibinfo {author} {\bibfnamefont {V.}~\bibnamefont {Cardoso}}, \ and\
  \bibinfo {author} {\bibfnamefont {M.}~\bibnamefont {Cavaglia}},\ }\href
  {\doibase 10.1103/PhysRevD.76.104044} {\bibfield  {journal} {\bibinfo
  {journal} {Phys. Rev. D}\ }\textbf {\bibinfo {volume} {76}},\ \bibinfo
  {pages} {104044} (\bibinfo {year} {2007})},\ \Eprint
  {http://arxiv.org/abs/0707.1202} {arXiv:0707.1202 [gr-qc]} \BibitemShut
  {NoStop}%
\bibitem [{\citenamefont {Baibhav}\ \emph {et~al.}(2018)\citenamefont
  {Baibhav}, \citenamefont {Berti}, \citenamefont {Cardoso},\ and\
  \citenamefont {Khanna}}]{Baibhav:2017jhs}%
  \BibitemOpen
  \bibfield  {author} {\bibinfo {author} {\bibfnamefont {V.}~\bibnamefont
  {Baibhav}}, \bibinfo {author} {\bibfnamefont {E.}~\bibnamefont {Berti}},
  \bibinfo {author} {\bibfnamefont {V.}~\bibnamefont {Cardoso}}, \ and\
  \bibinfo {author} {\bibfnamefont {G.}~\bibnamefont {Khanna}},\ }\href
  {\doibase 10.1103/PhysRevD.97.044048} {\bibfield  {journal} {\bibinfo
  {journal} {Phys. Rev. D}\ }\textbf {\bibinfo {volume} {97}},\ \bibinfo
  {pages} {044048} (\bibinfo {year} {2018})},\ \Eprint
  {http://arxiv.org/abs/1710.02156} {arXiv:1710.02156 [gr-qc]} \BibitemShut
  {NoStop}%
\bibitem [{\citenamefont {Abbott}\ \emph
  {et~al.}(2016{\natexlab{b}})\citenamefont {Abbott} \emph
  {et~al.}}]{LIGOScientific:2016lio}%
  \BibitemOpen
  \bibfield  {author} {\bibinfo {author} {\bibfnamefont {B.~P.}\ \bibnamefont
  {Abbott}} \emph {et~al.} (\bibinfo {collaboration} {LIGO Scientific,
  Virgo}),\ }\href {\doibase 10.1103/PhysRevLett.116.221101} {\bibfield
  {journal} {\bibinfo  {journal} {Phys. Rev. Lett.}\ }\textbf {\bibinfo
  {volume} {116}},\ \bibinfo {pages} {221101} (\bibinfo {year}
  {2016}{\natexlab{b}})},\ \bibinfo {note} {[Erratum: Phys.Rev.Lett. 121,
  129902 (2018)]},\ \Eprint {http://arxiv.org/abs/1602.03841} {arXiv:1602.03841
  [gr-qc]} \BibitemShut {NoStop}%
\bibitem [{\citenamefont {Isi}\ \emph {et~al.}(2019)\citenamefont {Isi},
  \citenamefont {Giesler}, \citenamefont {Farr}, \citenamefont {Scheel},\ and\
  \citenamefont {Teukolsky}}]{Isi:2019aib}%
  \BibitemOpen
  \bibfield  {author} {\bibinfo {author} {\bibfnamefont {M.}~\bibnamefont
  {Isi}}, \bibinfo {author} {\bibfnamefont {M.}~\bibnamefont {Giesler}},
  \bibinfo {author} {\bibfnamefont {W.~M.}\ \bibnamefont {Farr}}, \bibinfo
  {author} {\bibfnamefont {M.~A.}\ \bibnamefont {Scheel}}, \ and\ \bibinfo
  {author} {\bibfnamefont {S.~A.}\ \bibnamefont {Teukolsky}},\ }\href {\doibase
  10.1103/PhysRevLett.123.111102} {\bibfield  {journal} {\bibinfo  {journal}
  {Phys. Rev. Lett.}\ }\textbf {\bibinfo {volume} {123}},\ \bibinfo {pages}
  {111102} (\bibinfo {year} {2019})},\ \Eprint
  {http://arxiv.org/abs/1905.00869} {arXiv:1905.00869 [gr-qc]} \BibitemShut
  {NoStop}%
\bibitem [{\citenamefont {Carullo}\ \emph {et~al.}(2019)\citenamefont
  {Carullo}, \citenamefont {Del~Pozzo},\ and\ \citenamefont
  {Veitch}}]{Carullo:2019flw}%
  \BibitemOpen
  \bibfield  {author} {\bibinfo {author} {\bibfnamefont {G.}~\bibnamefont
  {Carullo}}, \bibinfo {author} {\bibfnamefont {W.}~\bibnamefont {Del~Pozzo}},
  \ and\ \bibinfo {author} {\bibfnamefont {J.}~\bibnamefont {Veitch}},\ }\href
  {\doibase 10.1103/PhysRevD.99.123029} {\bibfield  {journal} {\bibinfo
  {journal} {Phys. Rev. D}\ }\textbf {\bibinfo {volume} {99}},\ \bibinfo
  {pages} {123029} (\bibinfo {year} {2019})},\ \bibinfo {note} {[Erratum:
  Phys.Rev.D 100, 089903 (2019)]},\ \Eprint {http://arxiv.org/abs/1902.07527}
  {arXiv:1902.07527 [gr-qc]} \BibitemShut {NoStop}%
\bibitem [{\citenamefont {Laghi}\ \emph {et~al.}(2021)\citenamefont {Laghi},
  \citenamefont {Carullo}, \citenamefont {Veitch},\ and\ \citenamefont
  {Del~Pozzo}}]{Laghi:2020rgl}%
  \BibitemOpen
  \bibfield  {author} {\bibinfo {author} {\bibfnamefont {D.}~\bibnamefont
  {Laghi}}, \bibinfo {author} {\bibfnamefont {G.}~\bibnamefont {Carullo}},
  \bibinfo {author} {\bibfnamefont {J.}~\bibnamefont {Veitch}}, \ and\ \bibinfo
  {author} {\bibfnamefont {W.}~\bibnamefont {Del~Pozzo}},\ }\href {\doibase
  10.1088/1361-6382/abde19} {\bibfield  {journal} {\bibinfo  {journal} {Class.
  Quant. Grav.}\ }\textbf {\bibinfo {volume} {38}},\ \bibinfo {pages} {095005}
  (\bibinfo {year} {2021})},\ \Eprint {http://arxiv.org/abs/2011.03816}
  {arXiv:2011.03816 [gr-qc]} \BibitemShut {NoStop}%
\bibitem [{\citenamefont {Bustillo}\ \emph {et~al.}(2021)\citenamefont
  {Bustillo}, \citenamefont {Lasky},\ and\ \citenamefont
  {Thrane}}]{Bustillo:2020buq}%
  \BibitemOpen
  \bibfield  {author} {\bibinfo {author} {\bibfnamefont {J.~C.}\ \bibnamefont
  {Bustillo}}, \bibinfo {author} {\bibfnamefont {P.~D.}\ \bibnamefont {Lasky}},
  \ and\ \bibinfo {author} {\bibfnamefont {E.}~\bibnamefont {Thrane}},\ }\href
  {\doibase 10.1103/PhysRevD.103.024041} {\bibfield  {journal} {\bibinfo
  {journal} {Phys. Rev. D}\ }\textbf {\bibinfo {volume} {103}},\ \bibinfo
  {pages} {024041} (\bibinfo {year} {2021})},\ \Eprint
  {http://arxiv.org/abs/2010.01857} {arXiv:2010.01857 [gr-qc]} \BibitemShut
  {NoStop}%
\bibitem [{\citenamefont {Capano}\ \emph {et~al.}(2021)\citenamefont {Capano},
  \citenamefont {Cabero}, \citenamefont {Westerweck}, \citenamefont {Abedi},
  \citenamefont {Kastha}, \citenamefont {Nitz}, \citenamefont {Nielsen},\ and\
  \citenamefont {Krishnan}}]{Capano:2021etf}%
  \BibitemOpen
  \bibfield  {author} {\bibinfo {author} {\bibfnamefont {C.~D.}\ \bibnamefont
  {Capano}}, \bibinfo {author} {\bibfnamefont {M.}~\bibnamefont {Cabero}},
  \bibinfo {author} {\bibfnamefont {J.}~\bibnamefont {Westerweck}}, \bibinfo
  {author} {\bibfnamefont {J.}~\bibnamefont {Abedi}}, \bibinfo {author}
  {\bibfnamefont {S.}~\bibnamefont {Kastha}}, \bibinfo {author} {\bibfnamefont
  {A.~H.}\ \bibnamefont {Nitz}}, \bibinfo {author} {\bibfnamefont {A.~B.}\
  \bibnamefont {Nielsen}}, \ and\ \bibinfo {author} {\bibfnamefont
  {B.}~\bibnamefont {Krishnan}},\ }\href@noop {} {\  (\bibinfo {year}
  {2021})},\ \Eprint {http://arxiv.org/abs/2105.05238} {arXiv:2105.05238
  [gr-qc]} \BibitemShut {NoStop}%
\bibitem [{\citenamefont {Isi}\ and\ \citenamefont {Farr}(2021)}]{Isi:2021iql}%
  \BibitemOpen
  \bibfield  {author} {\bibinfo {author} {\bibfnamefont {M.}~\bibnamefont
  {Isi}}\ and\ \bibinfo {author} {\bibfnamefont {W.~M.}\ \bibnamefont {Farr}},\
  }\href@noop {} {\  (\bibinfo {year} {2021})},\ \Eprint
  {http://arxiv.org/abs/2107.05609} {arXiv:2107.05609 [gr-qc]} \BibitemShut
  {NoStop}%
\bibitem [{\citenamefont {Leaver}(1986)}]{Leaver:1986gd}%
  \BibitemOpen
  \bibfield  {author} {\bibinfo {author} {\bibfnamefont {E.~W.}\ \bibnamefont
  {Leaver}},\ }\href {\doibase 10.1103/PhysRevD.34.384} {\bibfield  {journal}
  {\bibinfo  {journal} {Phys. Rev. D}\ }\textbf {\bibinfo {volume} {34}},\
  \bibinfo {pages} {384} (\bibinfo {year} {1986})}\BibitemShut {NoStop}%
\bibitem [{\citenamefont {Kokkotas}\ and\ \citenamefont
  {Schmidt}(1999)}]{Kokkotas:1999bd}%
  \BibitemOpen
  \bibfield  {author} {\bibinfo {author} {\bibfnamefont {K.~D.}\ \bibnamefont
  {Kokkotas}}\ and\ \bibinfo {author} {\bibfnamefont {B.~G.}\ \bibnamefont
  {Schmidt}},\ }\href {\doibase 10.12942/lrr-1999-2} {\bibfield  {journal}
  {\bibinfo  {journal} {Living Rev. Rel.}\ }\textbf {\bibinfo {volume} {2}},\
  \bibinfo {pages} {2} (\bibinfo {year} {1999})},\ \Eprint
  {http://arxiv.org/abs/gr-qc/9909058} {arXiv:gr-qc/9909058} \BibitemShut
  {NoStop}%
\bibitem [{\citenamefont {Berti}\ \emph {et~al.}(2009)\citenamefont {Berti},
  \citenamefont {Cardoso},\ and\ \citenamefont {Starinets}}]{Berti:2009kk}%
  \BibitemOpen
  \bibfield  {author} {\bibinfo {author} {\bibfnamefont {E.}~\bibnamefont
  {Berti}}, \bibinfo {author} {\bibfnamefont {V.}~\bibnamefont {Cardoso}}, \
  and\ \bibinfo {author} {\bibfnamefont {A.~O.}\ \bibnamefont {Starinets}},\
  }\href {\doibase 10.1088/0264-9381/26/16/163001} {\bibfield  {journal}
  {\bibinfo  {journal} {Class. Quant. Grav.}\ }\textbf {\bibinfo {volume}
  {26}},\ \bibinfo {pages} {163001} (\bibinfo {year} {2009})},\ \Eprint
  {http://arxiv.org/abs/0905.2975} {arXiv:0905.2975 [gr-qc]} \BibitemShut
  {NoStop}%
\bibitem [{\citenamefont {{Trefethen}}\ \emph {et~al.}(1993)\citenamefont
  {{Trefethen}}, \citenamefont {{Trefethen}}, \citenamefont {{Reddy}},\ and\
  \citenamefont {{Driscoll}}}]{1993Sci...261..578T}%
  \BibitemOpen
  \bibfield  {author} {\bibinfo {author} {\bibfnamefont {L.~N.}\ \bibnamefont
  {{Trefethen}}}, \bibinfo {author} {\bibfnamefont {A.~E.}\ \bibnamefont
  {{Trefethen}}}, \bibinfo {author} {\bibfnamefont {S.~C.}\ \bibnamefont
  {{Reddy}}}, \ and\ \bibinfo {author} {\bibfnamefont {T.~A.}\ \bibnamefont
  {{Driscoll}}},\ }\href {\doibase 10.1126/science.261.5121.578} {\bibfield
  {journal} {\bibinfo  {journal} {Science}\ }\textbf {\bibinfo {volume}
  {261}},\ \bibinfo {pages} {578} (\bibinfo {year} {1993})}\BibitemShut
  {NoStop}%
\bibitem [{\citenamefont {Jaramillo}\ \emph
  {et~al.}(2021{\natexlab{a}})\citenamefont {Jaramillo}, \citenamefont
  {Panosso~Macedo},\ and\ \citenamefont {Al~Sheikh}}]{Jaramillo:2020tuu}%
  \BibitemOpen
  \bibfield  {author} {\bibinfo {author} {\bibfnamefont {J.~L.}\ \bibnamefont
  {Jaramillo}}, \bibinfo {author} {\bibfnamefont {R.}~\bibnamefont
  {Panosso~Macedo}}, \ and\ \bibinfo {author} {\bibfnamefont {L.}~\bibnamefont
  {Al~Sheikh}},\ }\href {\doibase 10.1103/PhysRevX.11.031003} {\bibfield
  {journal} {\bibinfo  {journal} {Phys. Rev. X}\ }\textbf {\bibinfo {volume}
  {11}},\ \bibinfo {pages} {031003} (\bibinfo {year} {2021}{\natexlab{a}})},\
  \Eprint {http://arxiv.org/abs/2004.06434} {arXiv:2004.06434 [gr-qc]}
  \BibitemShut {NoStop}%
\bibitem [{\citenamefont {Jaramillo}\ \emph
  {et~al.}(2021{\natexlab{b}})\citenamefont {Jaramillo}, \citenamefont
  {Panosso~Macedo},\ and\ \citenamefont {Sheikh}}]{Jaramillo:2021tmt}%
  \BibitemOpen
  \bibfield  {author} {\bibinfo {author} {\bibfnamefont {J.~L.}\ \bibnamefont
  {Jaramillo}}, \bibinfo {author} {\bibfnamefont {R.}~\bibnamefont
  {Panosso~Macedo}}, \ and\ \bibinfo {author} {\bibfnamefont {L.~A.}\
  \bibnamefont {Sheikh}},\ }\href@noop {} {\  (\bibinfo {year}
  {2021}{\natexlab{b}})},\ \Eprint {http://arxiv.org/abs/2105.03451}
  {arXiv:2105.03451 [gr-qc]} \BibitemShut {NoStop}%
\bibitem [{\citenamefont {Gasperin}\ and\ \citenamefont
  {Jaramillo}(2021)}]{Gasperin:2021kfv}%
  \BibitemOpen
  \bibfield  {author} {\bibinfo {author} {\bibfnamefont {E.}~\bibnamefont
  {Gasperin}}\ and\ \bibinfo {author} {\bibfnamefont {J.~L.}\ \bibnamefont
  {Jaramillo}},\ }\href@noop {} {\  (\bibinfo {year} {2021})},\ \Eprint
  {http://arxiv.org/abs/2107.12865} {arXiv:2107.12865 [gr-qc]} \BibitemShut
  {NoStop}%
\bibitem [{\citenamefont {Destounis}\ \emph {et~al.}(2021)\citenamefont
  {Destounis}, \citenamefont {Macedo}, \citenamefont {Berti}, \citenamefont
  {Cardoso},\ and\ \citenamefont {Jaramillo}}]{Destounis:2021lum}%
  \BibitemOpen
  \bibfield  {author} {\bibinfo {author} {\bibfnamefont {K.}~\bibnamefont
  {Destounis}}, \bibinfo {author} {\bibfnamefont {R.~P.}\ \bibnamefont
  {Macedo}}, \bibinfo {author} {\bibfnamefont {E.}~\bibnamefont {Berti}},
  \bibinfo {author} {\bibfnamefont {V.}~\bibnamefont {Cardoso}}, \ and\
  \bibinfo {author} {\bibfnamefont {J.~L.}\ \bibnamefont {Jaramillo}},\ }\href
  {\doibase 10.1103/PhysRevD.104.084091} {\bibfield  {journal} {\bibinfo
  {journal} {Phys. Rev. D}\ }\textbf {\bibinfo {volume} {104}},\ \bibinfo
  {pages} {084091} (\bibinfo {year} {2021})},\ \Eprint
  {http://arxiv.org/abs/2107.09673} {arXiv:2107.09673 [gr-qc]} \BibitemShut
  {NoStop}%
\bibitem [{\citenamefont {Nollert}(1996)}]{Nollert:1996rf}%
  \BibitemOpen
  \bibfield  {author} {\bibinfo {author} {\bibfnamefont {H.-P.}\ \bibnamefont
  {Nollert}},\ }\href {\doibase 10.1103/PhysRevD.53.4397} {\bibfield  {journal}
  {\bibinfo  {journal} {Phys. Rev. D}\ }\textbf {\bibinfo {volume} {53}},\
  \bibinfo {pages} {4397} (\bibinfo {year} {1996})},\ \Eprint
  {http://arxiv.org/abs/gr-qc/9602032} {arXiv:gr-qc/9602032} \BibitemShut
  {NoStop}%
\bibitem [{\citenamefont {Nollert}\ and\ \citenamefont
  {Price}(1999)}]{Nollert:1998ys}%
  \BibitemOpen
  \bibfield  {author} {\bibinfo {author} {\bibfnamefont {H.-P.}\ \bibnamefont
  {Nollert}}\ and\ \bibinfo {author} {\bibfnamefont {R.~H.}\ \bibnamefont
  {Price}},\ }\href {\doibase 10.1063/1.532698} {\bibfield  {journal} {\bibinfo
   {journal} {J. Math. Phys.}\ }\textbf {\bibinfo {volume} {40}},\ \bibinfo
  {pages} {980} (\bibinfo {year} {1999})},\ \Eprint
  {http://arxiv.org/abs/gr-qc/9810074} {arXiv:gr-qc/9810074} \BibitemShut
  {NoStop}%
\bibitem [{\citenamefont {Daghigh}\ \emph {et~al.}(2020)\citenamefont
  {Daghigh}, \citenamefont {Green},\ and\ \citenamefont
  {Morey}}]{Daghigh:2020jyk}%
  \BibitemOpen
  \bibfield  {author} {\bibinfo {author} {\bibfnamefont {R.~G.}\ \bibnamefont
  {Daghigh}}, \bibinfo {author} {\bibfnamefont {M.~D.}\ \bibnamefont {Green}},
  \ and\ \bibinfo {author} {\bibfnamefont {J.~C.}\ \bibnamefont {Morey}},\
  }\href {\doibase 10.1103/PhysRevD.101.104009} {\bibfield  {journal} {\bibinfo
   {journal} {Phys. Rev. D}\ }\textbf {\bibinfo {volume} {101}},\ \bibinfo
  {pages} {104009} (\bibinfo {year} {2020})},\ \Eprint
  {http://arxiv.org/abs/2002.07251} {arXiv:2002.07251 [gr-qc]} \BibitemShut
  {NoStop}%
\bibitem [{\citenamefont {Barausse}\ \emph {et~al.}(2014)\citenamefont
  {Barausse}, \citenamefont {Cardoso},\ and\ \citenamefont
  {Pani}}]{Barausse:2014tra}%
  \BibitemOpen
  \bibfield  {author} {\bibinfo {author} {\bibfnamefont {E.}~\bibnamefont
  {Barausse}}, \bibinfo {author} {\bibfnamefont {V.}~\bibnamefont {Cardoso}}, \
  and\ \bibinfo {author} {\bibfnamefont {P.}~\bibnamefont {Pani}},\ }\href
  {\doibase 10.1103/PhysRevD.89.104059} {\bibfield  {journal} {\bibinfo
  {journal} {Phys. Rev. D}\ }\textbf {\bibinfo {volume} {89}},\ \bibinfo
  {pages} {104059} (\bibinfo {year} {2014})},\ \Eprint
  {http://arxiv.org/abs/1404.7149} {arXiv:1404.7149 [gr-qc]} \BibitemShut
  {NoStop}%
\bibitem [{\citenamefont {Regge}\ and\ \citenamefont
  {Wheeler}(1957)}]{Regge:1957td}%
  \BibitemOpen
  \bibfield  {author} {\bibinfo {author} {\bibfnamefont {T.}~\bibnamefont
  {Regge}}\ and\ \bibinfo {author} {\bibfnamefont {J.~A.}\ \bibnamefont
  {Wheeler}},\ }\href {\doibase 10.1103/PhysRev.108.1063} {\bibfield  {journal}
  {\bibinfo  {journal} {Phys. Rev.}\ }\textbf {\bibinfo {volume} {108}},\
  \bibinfo {pages} {1063} (\bibinfo {year} {1957})}\BibitemShut {NoStop}%
\bibitem [{\citenamefont {Chandrasekhar}(1985)}]{Chandrasekhar:1985kt}%
  \BibitemOpen
  \bibfield  {author} {\bibinfo {author} {\bibfnamefont {S.}~\bibnamefont
  {Chandrasekhar}},\ }\href@noop {} {\emph {\bibinfo {title} {{The mathematical
  theory of black holes}}}}\ (\bibinfo {year} {1985})\BibitemShut {NoStop}%
\bibitem [{\citenamefont {Chung}\ \emph {et~al.}(2021)\citenamefont {Chung},
  \citenamefont {Gais}, \citenamefont {Cheung},\ and\ \citenamefont
  {Li}}]{Chung:2021roh}%
  \BibitemOpen
  \bibfield  {author} {\bibinfo {author} {\bibfnamefont {A.~K.-W.}\
  \bibnamefont {Chung}}, \bibinfo {author} {\bibfnamefont {J.}~\bibnamefont
  {Gais}}, \bibinfo {author} {\bibfnamefont {M.~H.-Y.}\ \bibnamefont {Cheung}},
  \ and\ \bibinfo {author} {\bibfnamefont {T.~G.~F.}\ \bibnamefont {Li}},\
  }\href {\doibase 10.1103/PhysRevD.104.084028} {\bibfield  {journal} {\bibinfo
   {journal} {Phys. Rev. D}\ }\textbf {\bibinfo {volume} {104}},\ \bibinfo
  {pages} {084028} (\bibinfo {year} {2021})},\ \Eprint
  {http://arxiv.org/abs/2107.05492} {arXiv:2107.05492 [gr-qc]} \BibitemShut
  {NoStop}%
\bibitem [{\citenamefont {Pani}(2013)}]{Pani:2013pma}%
  \BibitemOpen
  \bibfield  {author} {\bibinfo {author} {\bibfnamefont {P.}~\bibnamefont
  {Pani}},\ }\href {\doibase 10.1142/S0217751X13400186} {\bibfield  {journal}
  {\bibinfo  {journal} {Int. J. Mod. Phys. A}\ }\textbf {\bibinfo {volume}
  {28}},\ \bibinfo {pages} {1340018} (\bibinfo {year} {2013})},\ \Eprint
  {http://arxiv.org/abs/1305.6759} {arXiv:1305.6759 [gr-qc]} \BibitemShut
  {NoStop}%
\bibitem [{\citenamefont {Ansorg}\ and\ \citenamefont
  {Panosso~Macedo}(2016)}]{Ansorg:2016ztf}%
  \BibitemOpen
  \bibfield  {author} {\bibinfo {author} {\bibfnamefont {M.}~\bibnamefont
  {Ansorg}}\ and\ \bibinfo {author} {\bibfnamefont {R.}~\bibnamefont
  {Panosso~Macedo}},\ }\href {\doibase 10.1103/PhysRevD.93.124016} {\bibfield
  {journal} {\bibinfo  {journal} {Phys. Rev. D}\ }\textbf {\bibinfo {volume}
  {93}},\ \bibinfo {pages} {124016} (\bibinfo {year} {2016})},\ \Eprint
  {http://arxiv.org/abs/1604.02261} {arXiv:1604.02261 [gr-qc]} \BibitemShut
  {NoStop}%
\bibitem [{\citenamefont {Cardoso}\ \emph
  {et~al.}(2018{\natexlab{a}})\citenamefont {Cardoso}, \citenamefont {Costa},
  \citenamefont {Destounis}, \citenamefont {Hintz},\ and\ \citenamefont
  {Jansen}}]{Cardoso:2017soq}%
  \BibitemOpen
  \bibfield  {author} {\bibinfo {author} {\bibfnamefont {V.}~\bibnamefont
  {Cardoso}}, \bibinfo {author} {\bibfnamefont {J.~a.~L.}\ \bibnamefont
  {Costa}}, \bibinfo {author} {\bibfnamefont {K.}~\bibnamefont {Destounis}},
  \bibinfo {author} {\bibfnamefont {P.}~\bibnamefont {Hintz}}, \ and\ \bibinfo
  {author} {\bibfnamefont {A.}~\bibnamefont {Jansen}},\ }\href {\doibase
  10.1103/PhysRevLett.120.031103} {\bibfield  {journal} {\bibinfo  {journal}
  {Phys. Rev. Lett.}\ }\textbf {\bibinfo {volume} {120}},\ \bibinfo {pages}
  {031103} (\bibinfo {year} {2018}{\natexlab{a}})},\ \Eprint
  {http://arxiv.org/abs/1711.10502} {arXiv:1711.10502 [gr-qc]} \BibitemShut
  {NoStop}%
\bibitem [{\citenamefont {{Jona-Lasinio}}\ \emph {et~al.}(1981)\citenamefont
  {{Jona-Lasinio}}, \citenamefont {{Martinelli}},\ and\ \citenamefont
  {{Scoppola}}}]{1981CMaPh..80..223J}%
  \BibitemOpen
  \bibfield  {author} {\bibinfo {author} {\bibfnamefont {G.}~\bibnamefont
  {{Jona-Lasinio}}}, \bibinfo {author} {\bibfnamefont {F.}~\bibnamefont
  {{Martinelli}}}, \ and\ \bibinfo {author} {\bibfnamefont {E.}~\bibnamefont
  {{Scoppola}}},\ }\href {\doibase 10.1007/BF01213012} {\bibfield  {journal}
  {\bibinfo  {journal} {Communications in Mathematical Physics}\ }\textbf
  {\bibinfo {volume} {80}},\ \bibinfo {pages} {223} (\bibinfo {year}
  {1981})}\BibitemShut {NoStop}%
\bibitem [{\citenamefont {{Graffi}}\ \emph {et~al.}(1984)\citenamefont
  {{Graffi}}, \citenamefont {{Grecchi}},\ and\ \citenamefont
  {{Jona-Lasinio}}}]{1984JPhA...17.2935G}%
  \BibitemOpen
  \bibfield  {author} {\bibinfo {author} {\bibfnamefont {S.}~\bibnamefont
  {{Graffi}}}, \bibinfo {author} {\bibfnamefont {V.}~\bibnamefont {{Grecchi}}},
  \ and\ \bibinfo {author} {\bibfnamefont {G.}~\bibnamefont {{Jona-Lasinio}}},\
  }\href {\doibase 10.1088/0305-4470/17/15/011} {\bibfield  {journal} {\bibinfo
   {journal} {Journal of Physics A Mathematical General}\ }\textbf {\bibinfo
  {volume} {17}},\ \bibinfo {pages} {2935} (\bibinfo {year}
  {1984})}\BibitemShut {NoStop}%
\bibitem [{\citenamefont {Simon}(1985)}]{Simon1985SemiclassicalAO}%
  \BibitemOpen
  \bibfield  {author} {\bibinfo {author} {\bibfnamefont {B.}~\bibnamefont
  {Simon}},\ }\href {\doibase https://doi.org/10.1016/0022-1236(85)90101-6}
  {\bibfield  {journal} {\bibinfo  {journal} {Journal of Functional Analysis}\
  }\textbf {\bibinfo {volume} {63}},\ \bibinfo {pages} {123} (\bibinfo {year}
  {1985})}\BibitemShut {NoStop}%
\bibitem [{\citenamefont {{Santarsiero}}\ and\ \citenamefont
  {{Gori}}(2019)}]{2019EJPh...40e5402S}%
  \BibitemOpen
  \bibfield  {author} {\bibinfo {author} {\bibfnamefont {M.}~\bibnamefont
  {{Santarsiero}}}\ and\ \bibinfo {author} {\bibfnamefont {F.}~\bibnamefont
  {{Gori}}},\ }\href {\doibase 10.1088/1361-6404/ab2e6b} {\bibfield  {journal}
  {\bibinfo  {journal} {European Journal of Physics}\ }\textbf {\bibinfo
  {volume} {40}},\ \bibinfo {pages} {055402} (\bibinfo {year}
  {2019})}\BibitemShut {NoStop}%
\bibitem [{\citenamefont {Cardoso}\ \emph {et~al.}(2016)\citenamefont
  {Cardoso}, \citenamefont {Franzin},\ and\ \citenamefont
  {Pani}}]{Cardoso:2016rao}%
  \BibitemOpen
  \bibfield  {author} {\bibinfo {author} {\bibfnamefont {V.}~\bibnamefont
  {Cardoso}}, \bibinfo {author} {\bibfnamefont {E.}~\bibnamefont {Franzin}}, \
  and\ \bibinfo {author} {\bibfnamefont {P.}~\bibnamefont {Pani}},\ }\href
  {\doibase 10.1103/PhysRevLett.116.171101} {\bibfield  {journal} {\bibinfo
  {journal} {Phys. Rev. Lett.}\ }\textbf {\bibinfo {volume} {116}},\ \bibinfo
  {pages} {171101} (\bibinfo {year} {2016})},\ \bibinfo {note} {[Erratum:
  Phys.Rev.Lett. 117, 089902 (2016)]},\ \Eprint
  {http://arxiv.org/abs/1602.07309} {arXiv:1602.07309 [gr-qc]} \BibitemShut
  {NoStop}%
\bibitem [{\citenamefont {Yang}\ \emph {et~al.}(2015)\citenamefont {Yang},
  \citenamefont {Zimmerman},\ and\ \citenamefont {Lehner}}]{Yang:2014tla}%
  \BibitemOpen
  \bibfield  {author} {\bibinfo {author} {\bibfnamefont {H.}~\bibnamefont
  {Yang}}, \bibinfo {author} {\bibfnamefont {A.}~\bibnamefont {Zimmerman}}, \
  and\ \bibinfo {author} {\bibfnamefont {L.}~\bibnamefont {Lehner}},\ }\href
  {\doibase 10.1103/PhysRevLett.114.081101} {\bibfield  {journal} {\bibinfo
  {journal} {Phys. Rev. Lett.}\ }\textbf {\bibinfo {volume} {114}},\ \bibinfo
  {pages} {081101} (\bibinfo {year} {2015})},\ \Eprint
  {http://arxiv.org/abs/1402.4859} {arXiv:1402.4859 [gr-qc]} \BibitemShut
  {NoStop}%
\bibitem [{\citenamefont {Cardoso}\ \emph
  {et~al.}(2018{\natexlab{b}})\citenamefont {Cardoso}, \citenamefont {Costa},
  \citenamefont {Destounis}, \citenamefont {Hintz},\ and\ \citenamefont
  {Jansen}}]{Cardoso:2018nvb}%
  \BibitemOpen
  \bibfield  {author} {\bibinfo {author} {\bibfnamefont {V.}~\bibnamefont
  {Cardoso}}, \bibinfo {author} {\bibfnamefont {J.~L.}\ \bibnamefont {Costa}},
  \bibinfo {author} {\bibfnamefont {K.}~\bibnamefont {Destounis}}, \bibinfo
  {author} {\bibfnamefont {P.}~\bibnamefont {Hintz}}, \ and\ \bibinfo {author}
  {\bibfnamefont {A.}~\bibnamefont {Jansen}},\ }\href {\doibase
  10.1103/PhysRevD.98.104007} {\bibfield  {journal} {\bibinfo  {journal} {Phys.
  Rev. D}\ }\textbf {\bibinfo {volume} {98}},\ \bibinfo {pages} {104007}
  (\bibinfo {year} {2018}{\natexlab{b}})},\ \Eprint
  {http://arxiv.org/abs/1808.03631} {arXiv:1808.03631 [gr-qc]} \BibitemShut
  {NoStop}%
\bibitem [{\citenamefont {Cardoso}\ \emph {et~al.}(2021)\citenamefont
  {Cardoso}, \citenamefont {Destounis}, \citenamefont {Duque}, \citenamefont
  {Macedo},\ and\ \citenamefont {Maselli}}]{Cardoso:2021wlq}%
  \BibitemOpen
  \bibfield  {author} {\bibinfo {author} {\bibfnamefont {V.}~\bibnamefont
  {Cardoso}}, \bibinfo {author} {\bibfnamefont {K.}~\bibnamefont {Destounis}},
  \bibinfo {author} {\bibfnamefont {F.}~\bibnamefont {Duque}}, \bibinfo
  {author} {\bibfnamefont {R.~P.}\ \bibnamefont {Macedo}}, \ and\ \bibinfo
  {author} {\bibfnamefont {A.}~\bibnamefont {Maselli}},\ }\href@noop {} {\
  (\bibinfo {year} {2021})},\ \Eprint {http://arxiv.org/abs/2109.00005}
  {arXiv:2109.00005 [gr-qc]} \BibitemShut {NoStop}%
\bibitem [{\citenamefont {Chandrasekhar}\ and\ \citenamefont
  {Detweiler}(1975)}]{Chandrasekhar:1975zza}%
  \BibitemOpen
  \bibfield  {author} {\bibinfo {author} {\bibfnamefont {S.}~\bibnamefont
  {Chandrasekhar}}\ and\ \bibinfo {author} {\bibfnamefont {S.~L.}\ \bibnamefont
  {Detweiler}},\ }\href {\doibase 10.1098/rspa.1975.0112} {\bibfield  {journal}
  {\bibinfo  {journal} {Proc. Roy. Soc. Lond. A}\ }\textbf {\bibinfo {volume}
  {344}},\ \bibinfo {pages} {441} (\bibinfo {year} {1975})}\BibitemShut
  {NoStop}%
\end{thebibliography}%

\clearpage

\appendix
\section*{Supplemental material}

\begin{figure}[h]
	\includegraphics[width=0.48\textwidth]{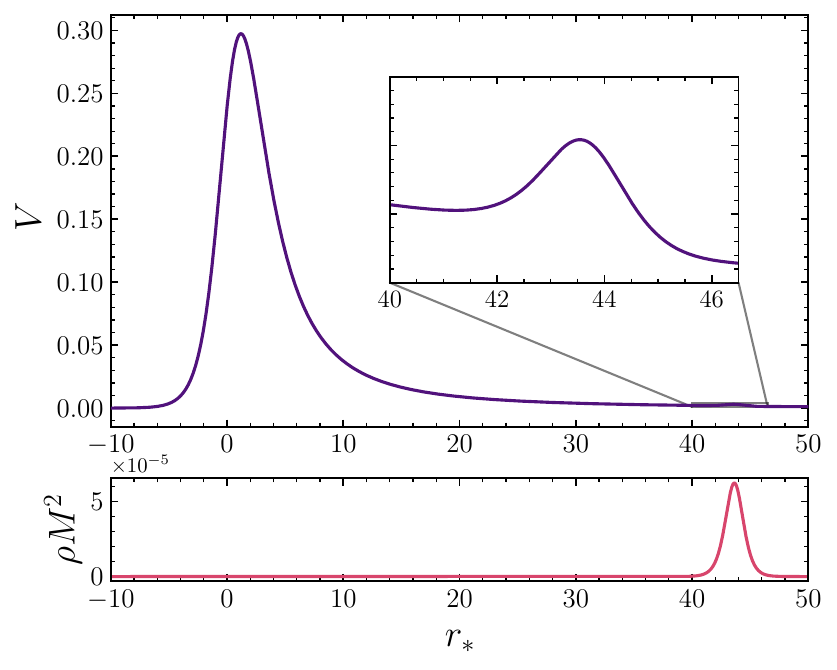}
	\caption{\label{fig:GR} \textit{Top:} Modified Regge-Wheeler potential due to a spherical matter shell with P\"oschl-Teller-like radial dependence.
	Here we used $l=2$, $\epsilon_m = 5$, $a_m = 40$ and $M = 1/2$.
	\textit{Bottom:} The density profile of the spherical matter shell. 
	The shell is localized at $r= 40$ ($r_* \sim 42$).}
\end{figure}

\noindent {\bf {\em Potential bump due to a matter shell.}}  Let us show that a spherical matter shell surrounding a BH could give rise to the perturbative bumps in the potential discussed in the main text.  Consider a spherical matter shell with density $\rho(r)$ having a P\"oschl-Teller-like peak as a function of the radial coordinate $r$:
\begin{equation}
	4 \pi r^2 \rho(r) = \epsilon_m \sech^2 (r - a_m),
\end{equation} 
where $\epsilon_m$ is proportional to the mass of the shell, and $a_m$ is the radial location of the peak of the shell.
We can integrate to find the mass $m(r)$ enclosed within the radial coordinate $r$, keeping in mind that the mass of the BH without the shell is $M$:
\begin{equation}
	m(r) = M + \epsilon_m [ 1 + \tanh (r - a_m) ].
\end{equation}
Then, the metric of such a spacetime will be given by
\begin{equation}
	ds^2 = -f(r) dt^2 + \dfrac{dr^2}{1-2m(r)/r} + r^2 d\Omega^2.
\end{equation}
Using the Einstein field equations, $f(r)$ can be determined numerically by solving the following ordinary differential equation:
\begin{equation}
\dfrac{r f^\prime(r)}{2 f(r)} = \dfrac{m(r)}{r - 2 m(r)}.
\end{equation} 
The matter shell will modify the Regge-Wheeler potential into 
\begin{equation}
V(r) = f(r) \left( \dfrac{l(l+1)}{r^2} + \dfrac{m^\prime(r)}{r^2} - \dfrac{6 m(r)}{r^3}\right).
\end{equation}
Fig.~\ref{fig:GR} shows $V(r)$ for the case $l=2$, $\epsilon_m = 5$, $a_m = 40$ and $M = 1/2$.
A bump at the location $r_*(r = 40) \sim 42$ is clearly visible.
We expect that, in general, localized spherical matter shells will introduce similar bumps in the Regge-Wheeler potential, possibly causing spectral instabilities.

\begin{figure}[th]
	\includegraphics[width=0.48\textwidth]{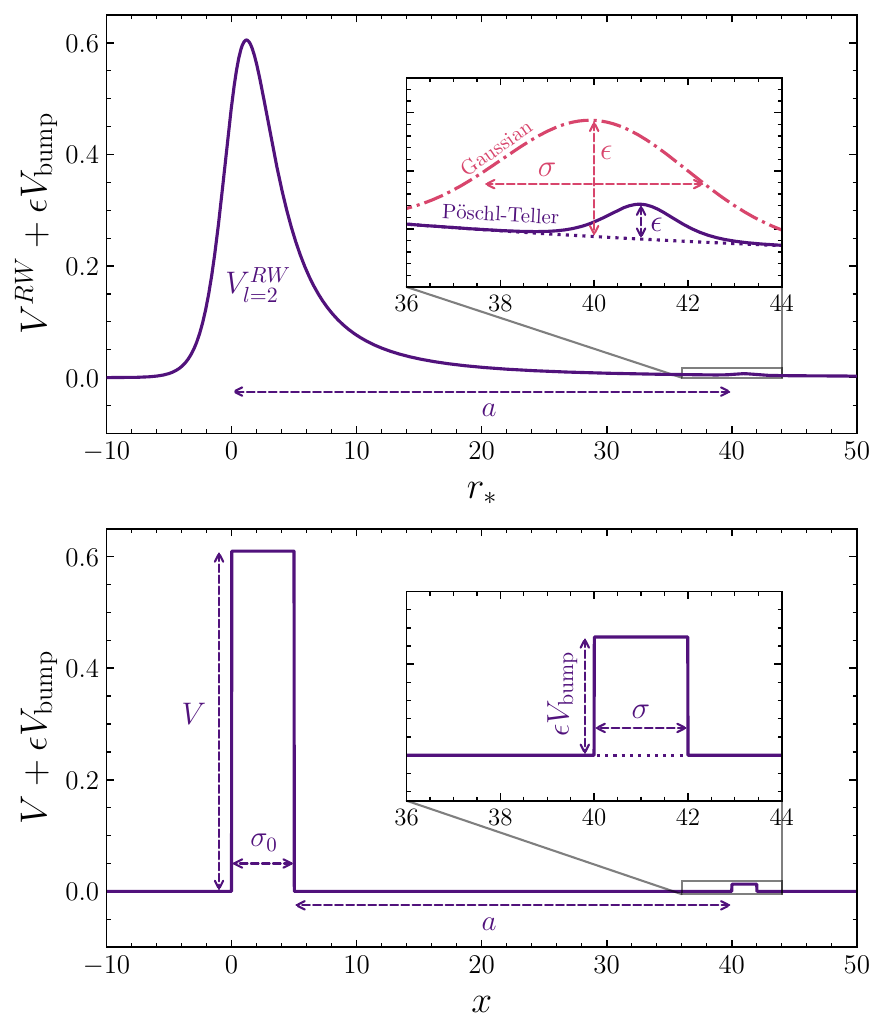}
	\caption{\label{fig:potentials} \textit{Top:} Schematic illustration of the Regge-Wheeler potential perturbed by a P\"oschl-Teller bump (solid line in the inset) and by a Gaussian bump (dot-dashed line).
	\textit{Bottom:} The double rectangular barrier potential toy model.}
\end{figure}

\begin{figure*}[th]
	\includegraphics[width=0.98\textwidth]{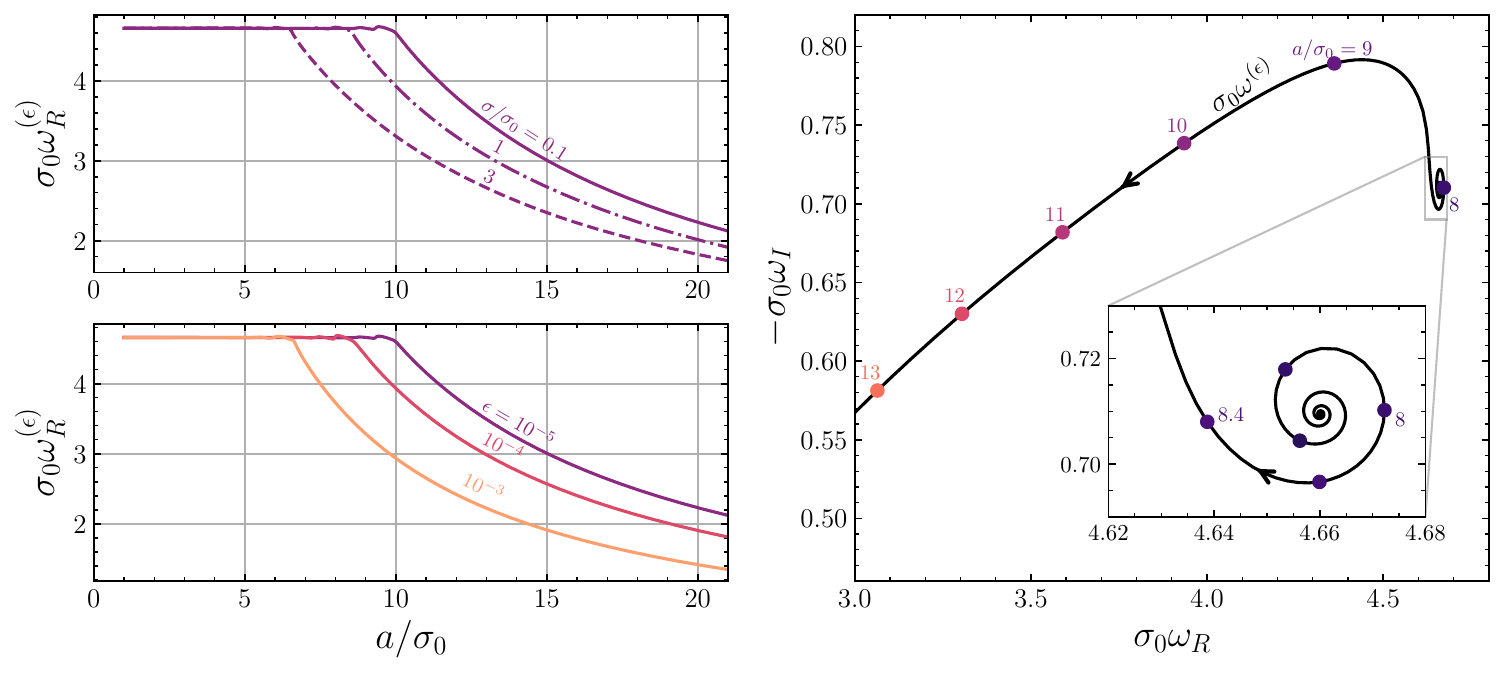}
	\caption{\label{fig:toy_model} \textit{Top left:} Real part of the fundamental QNM $\sigma_0 \omega_R^{(\epsilon)}$ of the double rectangular barrier model with $V=16/\sigma_0^2$, $\epsilon V_{\rm bump}=10^{-5}/\sigma^2_0$ as a function of the barrier's distance $a/\sigma_0$, for three selected values of $\sigma$: broader bumps destabilize the mode at smaller values of $a$. \textit{Bottom left:} We now fix $V=16/\sigma_0^2$, $V_{\rm bump}=1/\sigma^2_0$ and $\sigma/\sigma_0=0.1$, and we consider three selected values of the bump's amplitude $\epsilon$. As expected, larger values of $\epsilon$ destabilize the mode at smaller values of $a$. \textit{Right panel:} The fundamental QNM $\sigma_0 \omega^{(\epsilon)}$ of the double rectangular barrier model with $V=16/\sigma_0^2$, $\epsilon V_{\rm bump}=10^{-5}/\sigma^2_0$ and $\sigma/\sigma_0=1$ as a function of $a/\sigma_0$. The arrows indicate the direction of migration in the complex plane as $a$ increases.}
\end{figure*}

\noindent
{\bf {\em Double rectangular barrier potential: a toy model.}}
It is useful to consider a toy model which reproduces many features of the more realistic problem discussed in the main text. 
In Eq.~\eqref{eq:RW} we replace the perturbed Schwarzschild potential by a potential consisting of two rectangular barriers (see Fig. \ref{fig:potentials}): 
the barrier on the left mimics the potential barrier at $r\sim 3M$ in the Schwarzschild geometry, with height $V$ and width $\sigma_0$; the barrier on the right represents a ``bump'' with height $\epsilon V_{\rm bump}\ll V$ and width $\sigma$ at distance $a$ from the main barrier, as in Eq.~\eqref{eq:replacement}. When $\epsilon=0$, this model reduces to the single rectangular barrier considered in Ref.~\cite{Chandrasekhar:1975zza}.

A similar double rectangular barrier model was studied in Ref.~\cite{Barausse:2014tra}. Here we summarize the key features of the resulting QNM spectrum. 
We define QNMs as solutions which are left-moving at $x\rightarrow-\infty$ and right-moving at $x\rightarrow\infty$ (these correspond to ingoing waves at the horizon and outgoing waves at infinity). Then we can write
\begin{equation}
	\Psi_{\rm left}=\left\{\begin{array}{l}
		e^{-i\omega x} \qquad\qquad\qquad\qquad\quad x\leq 0 \\
		A_{\rm in}e^{-i k x}+A_{\rm out}e^{ik x} \quad 0\leq x\leq\, \sigma_0 \\
		B_{\rm in}e^{-i \omega x}+B_{\rm out}e^{i\omega x} \quad \sigma_0\leq x\leq\, a \\
		C_{\rm in}e^{-i k_{\epsilon} x}+C_{\rm out}e^{ik_{\epsilon} x} \quad a\leq x\leq\, \sigma \\
		D_{\rm in}e^{-i \omega x}+D_{\rm out}e^{i\omega x} \,\qquad x\geq \sigma\, 
	\end{array}\right.\,,\label{Qleft}
\end{equation}
where $k^2=\omega^2-V$, $k_\epsilon^2=\omega^2-\epsilon V_{\rm bump}$, and the coefficients $A, B, C, D$ can be found by requiring continuity of $\Psi$ and its derivative at the junctions $x=(0,\sigma_0,a,\sigma)$. 
A purely right-moving solution for $x\geq \sigma$ (i.e., a QNM) is found by requiring $D_{\rm out}(\omega)=0$.

In the left panel of Fig.~\ref{fig:toy_model} we plot the fundamental QNM frequency of the double-barrier potential as a function of the distance $a$ between the barriers.
The QNM frequency remains approximately constant up to a certain critical distance $a$ beyond which they undergo an outspiral motion (the ``migration instability'' of the main text), and subsequently new modes take the place of the original fundamental mode (``overtaking instability''). 
In the toy model, the fundamental QNM is destabilized most easily when the secondary bump is taller and wider: under these conditions, the fundamental QNM can be overtaken for smaller values of $a$. These qualitative conclusions are in complete agreement with the results presented in the main text.

\end{document}